\documentclass[twocolumn,showpacs,superscriptaddress,amsmath,amssymb,aps,pra]{revtex4-2}
\usepackage{graphicx}
\usepackage{CJKutf8}
\usepackage{dcolumn}
\usepackage{amsmath,bm}
\usepackage{epstopdf}
\usepackage[mathlines]{lineno}
\usepackage{color}
\usepackage[colorlinks=true,linkcolor=blue,citecolor=magenta,urlcolor=cyan]{hyperref}

\usepackage{tikz,xcolor,hyperref}

\definecolor{lime}{HTML}{A6CE39}
\DeclareRobustCommand{\orcidicon}{%
	\begin{tikzpicture}
	\draw[lime, fill=lime] (0,0) 
	circle [radius=0.16] 
	node[white] {{\fontfamily{qag}\selectfont \tiny ID}};
	\draw[white, fill=white] (-0.068,0.105) 
	circle [radius=0.007];
	\end{tikzpicture}
	\hspace{-2mm}
}

\foreach \x in {A, ..., Z}{%
	\expandafter\xdef\csname orcid\x\endcsname{\noexpand\href{https://orcid.org/\csname orcidauthor\x\endcsname}{\noexpand\orcidicon}}
}

\begin{document}
\title{Spin-photon interaction in a nanowire quantum dot with asymmetrical confining potential}
\author{Rui\! Li~(\begin{CJK}{UTF8}{gbsn}李睿\end{CJK})\orcidA{}}
\email{ruili@ysu.edu.cn}
\affiliation{Key Laboratory for Microstructural Material Physics of Hebei Province, School of Science, Yanshan University, Qinhuangdao 066004, China}

\begin{abstract}

The electron (hole) spin-photon interaction is studied in an asymmetrical InSb (Ge) nanowire quantum dot. The spin-orbit coupling in the quantum dot mediates not only a transverse spin-photon interaction, but also a longitudinal spin-photon interaction due to the asymmetry of  the confining potential. Both the transverse and the longitudinal spin-photon interactions have non-monotonic dependence on the spin-orbit coupling. For realistic spin-orbit coupling in the quantum dot, the longitudinal spin-photon interaction is much (at least one order) smaller than the transverse spin-photon interaction. The order of the transverse spin-photon interaction is about 1 nm in terms of length $|z_{eg}|$, or 0.1 MHz in terms of frequency $eE_{0}|z_{eg}|/h$ for a moderate cavity electric field strength $E_{0}=0.4$ V/m. 
\end{abstract}
\date{\today}
\maketitle

\section{Introduction}

The spin degree of freedom of electron or hole confined in semiconductor quantum dot can be utilized to design a qubit, the building block of a quantum computer~\cite{loss1998quantum,vandersypen2019,Scappucci:2021vk,RevModPhys.95.025003}. The single spin manipulation is achievable by using the electric-dipole spin resonance technique in quantum dot with spin-orbit coupling~\cite{PhysRevLett91126405,golovach2006electric,PhysRevLett.98.097202,RL2013,PhysRevB.92.054422,Wang:2021wc,PhysRevApplied.14.014090}, and the two-spin manipulation is achievable by using the exchange interaction in a double quantum dot~\cite{burkard1999coupled,hu2000hilbert}. Spin qubit also inevitably interacts with the surrounding electric~\cite{doi:10.1073/pnas.1603251113,yoneda2018}, magnetic~\cite{witzel2006quantum,yao2006theory}, and phonon environments~\cite{PhysRevB.61.12639,PhysRevB.64.125316}, such that there exist both spin relaxation and spin dephasing. The theory of lattice phonons induced spin relaxation in semiconductor quantum dot has been well established~\cite{PhysRevB.61.12639,PhysRevB.64.125316}. Also, several potential mechanisms of the charge noise induced spin dephasing were proposed recently~\cite{10.1063/1.4901162,RL2018c,RL2020}. 

The quantum theory of a two-level system interacting with cavity photons has been well built in quantum optics~\cite{scully1999quantum}. The basic models include the Jaynes-Cummings model~\cite{1443594} and the Rabi model~\cite{PhysRev.49.324}. There is an increasing interest in studying the spin-photon interaction in semiconductor quantum dot~\cite{trif2008spin,PhysRevB.86.035314,PhysRevB.88.241405,PhysRevB.102.205412,PhysRevLett.108.190506,Li:2018ab,Burkard:2020aa,PhysRevB.107.L041303,PhysRevLett.129.066801,PhysRevX.12.021026,PhysRevApplied.18.064082}, due to its potential application in achieving the long distance two spin interaction~\cite{PhysRevLett.83.4204,PhysRevLett.85.2392}. The spin-orbit coupling in semiconductor quantum dot provides a natural coupling mechanism for the spin and photons.  A one-dimensional (1D) cavity, i.e., a transmission line resonator~\cite{PhysRevA.69.062320,RevModPhys.93.025005}, is mainly used in current experiments~\cite{Petersson:2012aa,doi:10.1126/science.aar4054,Yu:2023aa,Bottcher:2022aa,PhysRevApplied.20.064005}. It is more convenient to achieve the strong system-photon interaction in a 1D cavity than in a traditional 3D cavity~\cite{PhysRevA.69.062320}. 


The quantum states of a nanowire quantum dot can be solved exactly even in the presence of strong spin-orbit coupling~\cite{RL2018b,RL2018c,RL2020}, such that it is possible to unveil novel spin-orbit coupling effect in this simple system. Previous studies obtained only a transverse spin-photon interaction in a nanowire quantum dot with symmetrical confining potential~\cite{trif2008spin,RL2013}. Here, we study the effects of the asymmetry of the confining potential~\cite{PhysRevA.91.023604} on the spin-photon interaction. We obtain exactly both the eigen-energies and the eigenstates of an asymmetrical quantum dot.  In the Hilbert subspace spanned by the lowest two eigenstates, we obtain a general spin-photon interaction Hamiltonian, in which in addition to the usual transverse spin-photon interaction~\cite{trif2008spin,PhysRevB.88.241405}, there is also a longitudinal spin-photon interaction. Both the transverse and the longitudinal spin-photon interactions have non-monotonic dependence on the spin-orbit coupling. The emergence of the longitudinal interaction is due to the combined effects of the spin-orbit coupling and the asymmetrical confining potential~\cite{RL2018c,RL2020}. For typical spin-orbit coupling and typical parameters of the quantum dot, the longitudinal interaction is at least one-order smaller than the transverse interaction. Although the longitudinal spin-electric interaction demonstrated here is too weak to affect the quantum information transfer between the spin and photons, it does provide a possible mechanism for the spin dephasing induced by the charge noise~\cite{RL2018c,RL2020}.

\section{Spin-photon interaction Hamiltonian}

\begin{figure}
\includegraphics{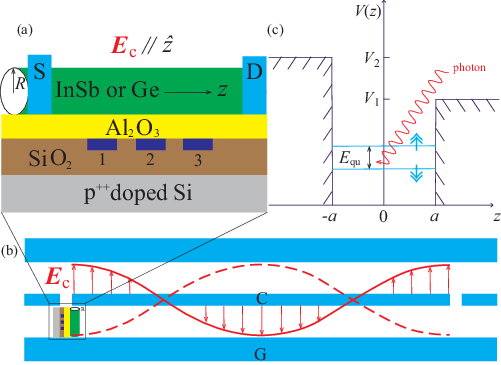}
\caption{\label{fig_qd}(a) A possible InSb or Ge nanowire quantum dot setup. The ${\rm p}^{++}$ doped Si substrate is covered with ${\rm SiO}_{2}$, three bottom gates 1-3 for creating the quantum dot confining potential are buried under ${\rm Al}_{2}{\rm O}_{3}$, on which an InSb or Ge nanowire is deposited. The contacts between the nanowire and source/drain are ohmic.  (b) The quantum dot setup is also placed in a transmission line resonator, and the cavity electric field ${\bf E}_{c}$ is along the nanowire axis, i.e., the $z$ direction. (c) An asymmetrical square well with heights $V_{1,2}$ and width $a$ is used to model the quantum dot confining potential. The spin basis states are just the lowest two eigenstates of the quantum dot.}
\end{figure}

The system under investigation is schematically shown in Fig.~\ref{fig_qd}. The undoped InSb (Ge) nanowire can be populated with electrons (holes) from the ohmic contacts via applying positive (negative) voltage to the backgate, i.e., the ${\rm p}^{++}$ doped Si substrate~\cite{nadj2012spectroscopy}. An electron (hole) can be trapped in the InSb~\cite{nadj2012spectroscopy,PhysRevLett.110.066806} (Ge)~\cite{Froning:2021aa,Wang:2022tm} nanowire quantum dot via applying proper voltages to the three bottom gates 1-3 [see Fig.~\ref{fig_qd}(a)]. The quantum dot setup is also placed in a 1D transmission line resonator~\cite{PhysRevA.69.062320,RevModPhys.93.025005,Petersson:2012aa,doi:10.1126/science.aar4054,Yu:2023aa,Bottcher:2022aa,PhysRevApplied.20.064005}, where the nanowire is placed to be perpendicular to the center conductor and pointing to the ground plane of the transmission line resonator [see Fig.~\ref{fig_qd}(b)]. The total Hamiltonian of this coupled quantum-dot-cavity system reads   
\begin{eqnarray}
H&=&\hbar^{2}k^{2}_{z}/(2m)+\alpha\sigma^{x}k_{z}+\delta\sigma^{z}+V(z)\nonumber\\
&&+ezE_{0}(b+b^{\dagger})+\hbar\omega_{c}\,b^{\dagger}b,\label{eq_Hamiltonian1}
\end{eqnarray}
where $m$ is the effective electron/hole mass, $\alpha$ is the spin-orbit coupling strength, $\delta$ is half of the Zeeman energy, $E_{0}$ is the strength of the cavity electric field ${\bf E}_{c}$, $\omega_{c}$ is the cavity frequency, $b$ ($b^{\dagger}$) is the bosonic annihilation (creation) operator for the cavity mode, and $V(z)$ is the  quantum dot confining potential modeled by an asymmetrical square-well [see Fig.~\ref{fig_qd}(c)]
\begin{equation}
V(z)=\left\{\begin{array}{cc}V_{2}, & z<-a,\\0, & |z|<a, \\V_{1}, & a<z,\end{array}\right.,\label{eq_potentialV}
\end{equation}
with $V_{1,2}$ and $a$ characterizing the well heights and the well width, respectively. Note that in our following calculations, unless otherwise stated, we have set the square well parameters as $a=40$ nm, $V_{1}=5$ meV, and $V_{2}=8$ meV. 

Our strategy of obtaining the effective spin-photon interaction Hamiltonian is as follows. First, we obtain exactly the basis states of the spin qubit, i.e., the lowest two eigenstates $|\Psi_{e,g}\rangle$ of the quantum dot Hamiltonian $H_{\rm qu}=\hbar^{2}k^{2}_{z}/(2m)+\alpha\sigma^{x}k_{z}+\delta\sigma^{z}+V(z)$, i.e., the first line of Eq.~(\ref{eq_Hamiltonian1}). Note that as long as $V(z)$ is a square well potential, the quantum dot Hamiltonian $H_{\rm qu}$ is exactly solvable~\cite{PhysRevB.91.235421,RL2018b,RL2018c,RL2020}, although $H_{\rm qu}$ contains non-trivial spin-orbit coupling physics. Second, in the Hilbert subspace spanned by the qubit basis states $|\Psi_{e,g}\rangle$, we calculate the matrix form of the electric-dipole operator $ez$ shown in Eq.~(\ref{eq_Hamiltonian1}). Finally, we are able to obtain an effective spin-photon interaction Hamiltonian $H_{\rm ef}$, that can be written explicitly as
\begin{eqnarray}
H_{\rm ef}&=&\frac{E_{\rm qu}}{2}\tilde{\sigma}^{z}+eE_{0}(b+b^{\dagger})\Big[\Big(\frac{z_{ee}+z_{gg}}{2}+\frac{z_{ee}-z_{gg}}{2}\tilde{\sigma}^{z}\Big)\nonumber\\
&&+|z_{eg}|(\tilde{\sigma}^{x}\cos\varphi_{eg}+\tilde{\sigma}^{y}\sin\varphi_{eg})\Big]+\hbar\omega_{c}b^{\dagger}b,\label{eq_spincavity}
\end{eqnarray}
where $E_{\rm qu}$ is the level splitting of the spin qubit,  $\tilde{\sigma}^{x,y,z}$ are the Pauli matrices defined in the qubit Hilbert subspace, $z_{ee}=\langle\Psi_{e}|z|\Psi_{e}\rangle$, $z_{gg}=\langle\Psi_{g}|z|\Psi_{g}\rangle$, and $z_{eg}=|z_{eg}|{\rm exp}(i\varphi_{eg})=\langle\Psi_{e}|z|\Psi_{g}\rangle$. Although the matrix element $z_{eg}$ has both amplitude $|z_{eg}|$ and phase $\varphi_{eg}$, the strength of the transverse interaction only depends on the amplitude $|z_{eg}|$.

The effective interaction Hamiltonian (\ref{eq_spincavity}), in which there are both longitudinal and transverse spin-photon interactions, can be regarded as a general spin-boson model in quantum optics. The strength of the spin-photon interaction is certainly proportional to the cavity electric field strength $E_{0}$. The realistic $E_{0}$ is related to the frequency of cavity and some other parameters of the transmission line~\cite{PhysRevA.69.062320}, and its typical value is about $E_{0}=0.4$ V/m~\cite{PhysRevA.69.062320}. However, this value may be improvable by future experimental advances, such that in the following we use $(z_{ee}-z_{gg})/2$ and $|z_{eg}|$ to represent the relative strength of the longitudinal and transverse spin-photon interactions, respectively. The unit of the absolute strength of the interactions can be regarded as $eE_{0}/h\approx0.1$ MHz/nm for $E_{0}=0.4$ V/m.

\section{Results in an InSb nanowire quantum dot}

In this section, we focus on the electron spin in an InSb nanowire quantum dot and its interaction with photons. Strong spin-orbit coupling is expected to exist in an InSb nanowire quantum dot, and a recent experiment has indeed reported a spin-orbit length of $z_{\rm so}=\hbar^{2}/(m\alpha)\approx230$ nm~\cite{nadj2012spectroscopy}. The spin manipulation frequency as large as 100 MHz by an external electric field has also been demonstrated~\cite{PhysRevLett.110.066806}. For an InSb nanowire quantum dot, the effective electron mass in Eq.~(\ref{eq_Hamiltonian1}) reads $m=0.0136m_{e}$~\cite{madelung2004semiconductors}, where $m_{e}$ is the free electron mass, and the parameters $\alpha$ and $\delta$ in Eq.~(\ref{eq_Hamiltonian1}) can be simply interpreted as the Rashba spin-orbit coupling~\cite{bychkov1984oscillatory} and half of the Zeeman energy in a magnetic field, respectively. We have
\begin{eqnarray}
\alpha&=&\alpha(E),\nonumber\\
\delta&=&g_{e}\mu_{B}B/2,\label{eq_parameters}
\end{eqnarray}
where the Rashba spin-orbit coupling may be tunable by an external static electric field $E$, and $g_{e}=-50.6$ is the electron g-factor in InSb~\cite{PhysRevB.70.115316}. Note that $\alpha$ may also contain a Dresselhaus spin-orbit coupling~\cite{PhysRev.100.580} contribution, this does not influence the form of Hamiltonian (\ref{eq_Hamiltonian1}), because the Zeeman field can always be tuned to be perpendicular to the spin-orbit field for quasi-1D case.

\begin{figure}
\includegraphics{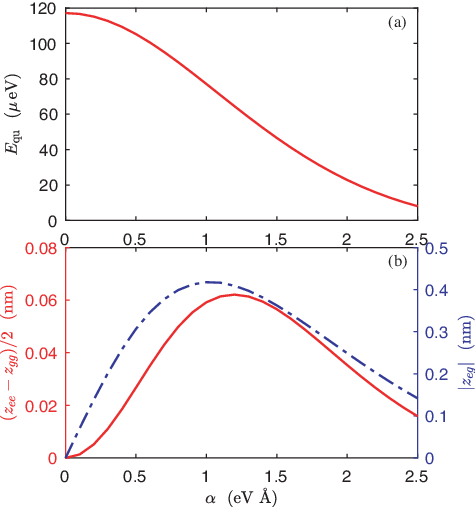}
\caption{\label{fig_InsbEvsSOC}(a) The spin splitting as a function of the spin-orbit coupling $\alpha$. (b) Both the longitudinal and the transverse spin-photon interactions as a function of the spin-orbit coupling $\alpha$. The magnetic field in the InSb quantum dot is chosen as $B=0.04$ T.}
\end{figure}

We show in Fig.~\ref{fig_InsbEvsSOC} the spin splitting $E_{\rm qu}$, the longitudinal spin-photon interaction $(z_{ee}-z_{gg})/2$, and the transverse spin-photon interaction $|z_{eg}|$ as a function of the spin-orbit coupling $\alpha$. Here, we have implicitly assumed $\alpha$ is tunable by an external static electric field, and the magnetic field in Eq.~(\ref{eq_parameters}) is fixed at $B=0.04$ T. With the increase of the spin-orbit coupling, the spin splitting $E_{\rm qu}$ decreases [see Fig.~\ref{fig_InsbEvsSOC}(a)], and both the longitudinal interaction $(z_{ee}-z_{gg})/2$ and the transverse interaction $|z_{eg}|$ first increase to a maximum and then decrease [see Fig.~\ref{fig_InsbEvsSOC}(b)]. Now, we can compare these results with that obtained in a symmetrical quantum dot. First, the spin-orbit coupling dependences of both $E_{\rm qu}\propto{\rm exp}(-z^{2}_{\rm size}/z^{2}_{\rm so})$ and $|z_{eg}|\propto\,z_{\rm size}/z_{\rm so}{\rm exp}(-z^{2}_{\rm size}/z^{2}_{\rm so})$ are similar to that in a symmetrical quantum dot~\cite{trif2008spin,PhysRevB.87.205436,RL2013}. Here, $z_{\rm size}$ and $z_{\rm so}=\hbar^{2}/(m\alpha)$ are the quantum dot size (or characteristic length) and the spin-orbit length, respectively. Second, the longitudinal interaction $(z_{ee}-z_{gg})/2$ can be observed only in an asymmetrical quantum dot.

\begin{figure}
\includegraphics{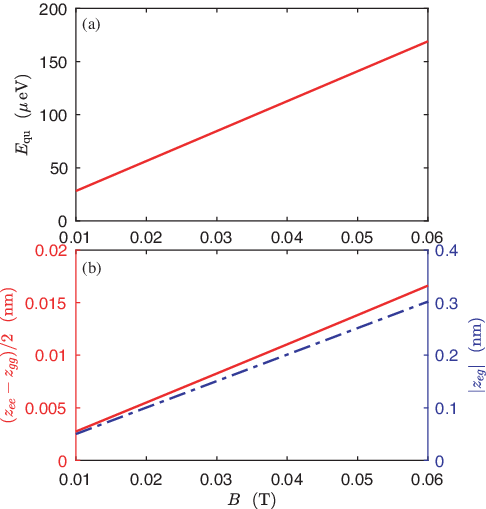}
\caption{\label{fig_InSbEvsB}(a) The spin splitting as a function of the magnetic field. (b) Both the longitudinal and the transverse spin-photon interactions as a function of the magnetic field. The spin-orbit coupling in the InSb quantum dot is chosen as $\alpha=0.3$ eV \AA~($z_{\rm so}=187$ nm).}
\end{figure}

In previous analytical studies in a symmetrical quantum dot, the Zeeman term $\delta$ [see Eq.~(\ref{eq_parameters})] is treated as a perturbation, such that all the physical quantities obtained have a linear magnetic field dependence~\cite{trif2008spin,PhysRevB.87.205436,RL2013}. Here, as expected, the spin splitting $E_{\rm qu}$, the longitudinal spin-photon interaction $(z_{ee}-z_{gg})/2$, and the transverse spin-photon interaction $|z_{eg}|$ all have a linear magnetic field dependence when the magnetic field is not very large (see Fig.~\ref{fig_InSbEvsB}). Here the spin-orbit coupling is set to a realistic value $\alpha=0.3$ eV~\AA, that corresponds to a spin-orbit length $z_{\rm so}=\hbar^{2}/(m\alpha)=187$ nm, comparable with the experimentally observed value 230 nm~\cite{nadj2012spectroscopy}. Note that for this realistic spin-orbit coupling, the longitudinal interaction is almost one-order smaller than the transverse interaction [see Fig.~\ref{fig_InSbEvsB}(b)]. 

\begin{figure}
\includegraphics{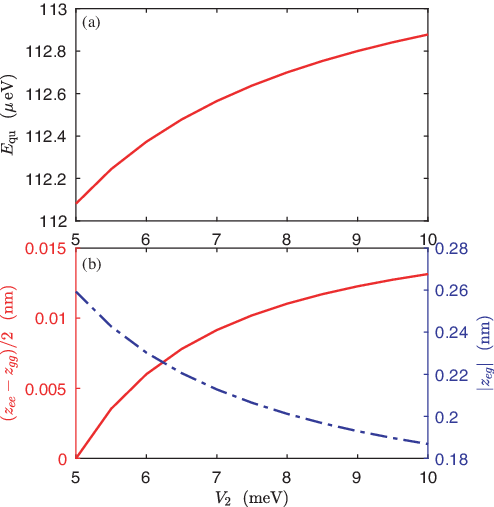}
\caption{\label{fig_InSbEvsV2}(a) The spin splitting as a function of the potential height $V_{2}$. (b) Both the longitudinal and the transverse spin-photon interactions as a function of the potential height $V_{2}$. Here the other potential height is fixed at $V_{1}=5$ meV, and the spin-orbit coupling and magnetic field in the InSb quantum dot are chosen as $\alpha=0.3$ eV \AA~($z_{\rm so}=187$ nm) and $B=0.04$ T, respectively.}
\end{figure}

As we have emphasized that the longitudinal spin-photon interaction exists only in an asymmetrical quantum dot, now this assertation is unambiguously demonstrated in Fig.~\ref{fig_InSbEvsV2}, where we show the spin splitting $E_{\rm qu}$, the longitudinal spin-photon interaction $(z_{ee}-z_{gg})/2$, and the transverse spin-photon interaction $|z_{eg}|$ as a function of the potential height $V_{2}$. When $V_{2}=5$ meV, the confining potential is symmetrical, and the longitudinal interaction is zero [see Fig.~\ref{fig_InSbEvsV2}(b)]. Here the spin-orbit coupling and magnetic field are chosen as $\alpha=0.3$ eV \AA~and $B=0.04$ T, respectively. In order to completely understand the results shown in Fig.~\ref{fig_InSbEvsV2}, we should know that with the increase of the potential height $V_{2}$, not only does the degree of the asymmetry of the confining potential increase, but also the quantum dot size $z_{\rm size}$ decreases slightly. For $\alpha=0.3$ eV~\AA, the spin-orbit length $z_{\rm so}=\hbar^{2}/(m\alpha)=187$ nm is much longer than the quantum dot size $z_{\rm size}\sim40$ nm. When $V_{2}$ increases, the quantum dot size decreases, such that $E_{\rm qu}\propto{\rm exp}(-z^{2}_{\rm size}/z^{2}_{\rm so})$ increases [see Fig.~\ref{fig_InSbEvsV2}(a)] and $|z_{eg}|\propto\,z_{\rm size}/z_{\rm so}{\rm exp}(-z^{2}_{\rm size}/z^{2}_{\rm so})\approx\,z_{\rm size}/z_{\rm so}$ decreases [see Fig.~\ref{fig_InSbEvsV2}(b)]~\cite{trif2008spin,RL2013}. While the longitudinal spin-photon interaction $(z_{ee}-z_{gg})/2$, resulting from the asymmetry of the confining potential, increases when $V_{2}$ increases [see Fig.~\ref{fig_InSbEvsV2}(b)]. Note that here the longitudinal interaction is still at least one-order smaller than the transverse interaction.

\section{Results in a Ge nanowire quantum dot}

\begin{figure}
\includegraphics{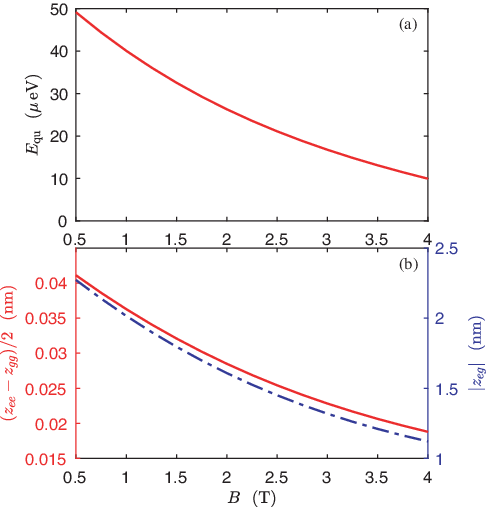}
\caption{\label{fig_GeEvsB}(a) The spin splitting as a function of the magnetic field. (b) Both the longitudinal and the transverse spin-photon interactions as a function of the magnetic field. }
\end{figure}

In this section, we focus on the hole pseudo spin in a Ge nanowire quantum dot and its interaction with photons. The hole in a cylindrical Ge nanowire has a peculiar low-energy subband structure, i.e., two shifted parabolic dispersions with an anticrossing at $k_{z}=0$~\cite{PhysRevB.84.195314,RL2022a,RL2023b,RL2023c}. This result indicates that there is a natural strong pseudo spin-orbit coupling for hole due to the subband quantization~\cite{PhysRevB.84.195314,RL2023b}, although the real spin degeneracy still exists in the subband dispersions. If we split out the real spin degree of freedom by using a strong magnetic field~\cite{RL2022a}, the induced lowest hole subband dispersion can be accurately described by the first three terms in Eq.~(\ref{eq_Hamiltonian1})~\cite{RL2023b}. The spin-orbit coupling obtained in this way has the same form as that obtained by using a strong electric field to induce the structure inversion asymmetry in the nanowire~\cite{PhysRevB.84.195314,PhysRevB.97.235422,PhysRevB.84.121303,PhysRevLett.119.126401}. The main difference of these two spin-orbit couplings is that, one is in terms of the pseudo spin and depends on the external magnetic field~\cite{RL2023b}, while the other is in terms of the real spin and depends on the external electric field~\cite{PhysRevB.84.195314,PhysRevLett.119.126401,PhysRevB.90.195421,PhysRevB.104.235304,PhysRevB.105.075308,PhysRevB.106.235408}.

The interaction Hamitloniain between a Ge nanowire quantum dot and the cavity can still be written as Eq.~(\ref{eq_Hamiltonian1}). Now, the effective hole mass in Ge reads $m\approx0.074m_{e}$~\cite{RL2023b}, and both the parameters $\alpha$ and $\delta$ have non-trivial magnetic field dependence~\cite{RL2023b}
\begin{eqnarray}
\alpha(B)&=&C\frac{\hbar^{2}}{m_{e}R}-\,Z_{3}\mu_{B}BR,\nonumber\\
\delta(B)&=&\frac{\Delta}{2}\frac{\hbar^{2}}{m_{e}R^{2}}-Z_{1}\mu_{B}B+Z_{4}\frac{e^{2}B^{2}R^{2}}{2m_{e}},\label{eq_parameters2}
\end{eqnarray}
where $C\approx7.12$, $\Delta\approx0.77$, $R$ is the radius of the nanowire,  $Z_{1}=0.75$, $Z_{3}=-3.62$, and $Z_{4}=0.65$~\cite{PhysRevB.84.195314,RL2023b}. Note that we have assumed the magnetic field is longitudinal, and the nanowire radius is set to $R=10$ nm in the following calculations. The spin-orbit length $z_{\rm so}=\hbar^{2}/(m\alpha)$ in this case is about $16\sim19$ nm depending on the magnetic field strength~\cite{RL2023b}, and it is less than the quantum dot width $a=40$ nm. 

We show in Fig.~\ref{fig_GeEvsB} the spin splitting $E_{\rm qu}$, the longitudinal spin-photon interaction $(z_{ee}-z_{gg})/2$, and the transverse spin-photon interaction $|z_{eg}|$ as a function of the longitudinal magnetic field $B$. With the increase of the magnetic field $B$, the spin-orbit coupling $\alpha(B)$ increases while the Zeeman energy $\delta(B)$ decreases [see Eq.~(\ref{eq_parameters2})], such that the spin splitting $E_{\rm qu}$ decreases reasonably [see Fig.~\ref{fig_GeEvsB}(a)]. The results shown in Fig.~\ref{fig_GeEvsB}(b) can be understood conveniently if we refer to the results shown in Fig.~\ref{fig_InsbEvsSOC}(b), that after the maximums of both $(z_{ee}-z_{gg})/2$ and $|z_{eg}|$, these two quantities will decrease with the increase of $\alpha$. Here the spin-orbit length $16\sim19$ nm in the Ge quantum dot is indeed smaller than the dot width $a=40$ nm, such that the situation here is indeed similar to that of the ultrastrong spin-orbit coupling region of the InSb quantum dot, e.g., $\alpha>1.2$ eV \AA~in Fig.~\ref{fig_InsbEvsSOC}(b). Therefore, with the increase of $B$, the spin-orbit coupling $\alpha$ increases, it is reasonable to see both the longitudinal interaction $(z_{ee}-z_{gg})/2$ and the transverse interaction $z_{eg}$ decrease [see Fig.~\ref{fig_GeEvsB}(b)]. We also note that the longitudinal interaction is at least one-order smaller than the transverse interaction. The order of the transverse interaction is 1 nm in terms of length $|z_{eg}|$, or 0.1 MHz in terms of frequency $eE_{0}|z_{eg}|/h$. 

\begin{figure}
\includegraphics{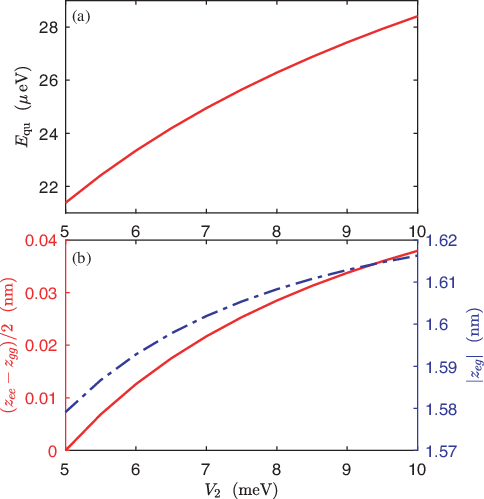}
\caption{\label{fig_GeEvsV2}(a) The spin splitting as a function of the potential height $V_{2}$. (b) Both the longitudinal and the transverse spin-photon interactions as a function of the potential height $V_{2}$. Here the other potential height is fixed at $V_{1}=5$ meV and the magnetic field in the Ge quantum dot is chosen as $B=2$ T.}
\end{figure}

In Fig.~\ref{fig_GeEvsV2}, we show the potential height $V_{2}$ dependence of the spin splitting $E_{\rm qu}$, the longitudinal spin-photon interaction $(z_{ee}-z_{gg})/2$, and the transverse spin-photon interaction $z_{eg}$. We should keep in mind that when we vary the potential height $V_{2}$, both the quantum dot size and the degree of asymmetry of the potential are varied. With the increase of $V_{2}$, the quantum dot $z_{\rm size}$ decreases slightly, such that the level splitting $E_{\rm qu}\propto{\rm exp}(-z^{2}_{\rm size}/z^{2}_{\rm so})$ increases [see Fig.~\ref{fig_GeEvsV2}(a)]~\cite{trif2008spin,PhysRevB.87.205436,RL2013}. Because the quantum dot is in the ultrastrong spin-orbit coupling region, that is similar to the case $\alpha>1.2$ eV \AA~in Fig.~\ref{fig_InsbEvsSOC}(b), it is not difficult to understand $|z_{eg}|\propto\,z_{\rm size}/z_{\rm so}{\rm exp}(-z^{2}_{\rm size}/z^{2}_{\rm so})$ increases with the increase of $V_{2}$ [see Fig.~\ref{fig_GeEvsV2}(b)]~\cite{trif2008spin,RL2013}. While for the longitudinal spin-photon interaction $(z_{ee}-z_{gg})/2$, resulting from the asymmetry of the confining potential, increases with the increase of $V_{2}$ [see Fig.~\ref{fig_GeEvsV2}(b)]. Also, when $V_{2}=5$ meV, the confining potential is symmetrical, there is no longitudinal spin-photon interaction. Note that here the longitudinal interaction is still at least one-order smaller than the transverse interaction.

\section{Charge noise induced spin dephasing}
In Eq.~(\ref{eq_spincavity}), $(z_{ee}+z_{gg})/2$ can be interpreted as the equilibrium point of the electron or hole in an asymmetrical confining potential. If we take this equilibrium point into account in the electric-dipole interaction $e[z-(z_{ee}+z_{gg})/2]E_{c}$, the $(z_{ee}+z_{gg})/2$ term in Eq.~(\ref{eq_spincavity}) will disappear. We now generalize Eq.~(\ref{eq_spincavity}) to the case of the spin-charge-noise interaction, where ${\bf E}_{c}$ is replaced by a fluctuating charge field $\sum_{k}E_{k}(b_{k}+b^{\dagger}_{k})$. The longitudinal spin-charge interaction induces spin dephasing ($T_{2}$), while the transverse spin-charge interaction induces spin relaxation ($T_{1}$). For quantum dot spin qubit, the dephasing time is usually much less than the relaxation time $T_{2}\ll\,T_{1}$~\cite{Buluta_2011}. 

Here we focus only on the dephasing time of the spin qubit, such that we have the following effective interaction Hamiltonian 
\begin{equation}
H_{\rm ef}=\frac{E_{\rm qu}}{2}\tilde{\sigma}^{z}+e\frac{z_{ee}-z_{gg}}{2}\tilde{\sigma}^{z}\sum_{k}E_{k}(b_{k}+b^{\dagger}_{k})+\sum_{k}\hbar\omega_{k}b^{\dagger}_{k}b_{k}.\label{eq_charge_noise}
\end{equation}
The off-diagonal element of the reduced density matrix of the spin qubit  governed by Hamiltonian~(\ref{eq_charge_noise}) can be solved as $\rho_{\uparrow\downarrow}(t)=\rho_{\uparrow\downarrow}(0){\rm exp}[-\Gamma(t)]$, where~\cite{RL2018c,RL2020}
\begin{equation}
\Gamma(t)=\sum_{k}\frac{2(z_{ee}-z_{gg})^{2}e^{2}E^{2}_{k}\sin^{2}\frac{\omega_{k}t}{2}}{\hbar^{2}\omega^{2}_{k}}[2n(\omega_{k})+1],\label{eq_dephase_factor}
\end{equation}
with $n(\omega_{k})=1/\left[{\rm exp}(\hbar\omega_{k}/k_{B}T)-1\right]$ being the thermal average boson number. 

In order to obtain the dephasing time of the spin qubit, one has to assume a strength of the spectrum function of the charge noise~\cite{RL2018c,RL2020}, a quantity related to the series of $E_{k}$ values in Eq.~(\ref{eq_charge_noise}). Here, we will not do this calculation by artificially choosing a strength of the spectrum function, we just compare the dephasing time of the spin qubit with that of the charge qubit in a double quantum dot. When the noise frequency spectrum has both lower bound $\omega_{\rm min}$ and upper bound $\omega_{\rm max}$ such as the $1/f$ charge noise~\cite{RevModPhys.86.361}, i.e., $\omega_{k}\in[\omega_{\rm min},\omega_{\rm max}]$, for time scales $t<1/\omega_{\rm max}$, we can safely make the following approximation $\sin^{2}\omega_{k}t/2\approx\omega^{2}_{k}t^{2}/4$ in Eq.~(\ref{eq_dephase_factor}), such that we have
\begin{equation}
\Gamma(t)\propto(z_{ee}-z_{gg})^{2}t^{2}.\label{eq_Gamma}
\end{equation}
For a charge qubit in double quantum dot, $(z_{ee}-z_{gg})_{\rm charge}$ has an order of magnitude of the interdot distance, the typical value is about several tens of nm, while for the spin qubit considered here, $(z_{ee}-z_{gg})_{\rm spin}$ has an order of magnitude of $10^{-2}$ nm. Thus, it is reasonable to have the following relation $(z_{ee}-z_{gg})_{\rm spin}\approx10^{-3}(z_{ee}-z_{gg})_{\rm charge}$. A rough analysis of Eq.~(\ref{eq_Gamma}) gives rise to $T^{\rm spin}_{2}\approx10^{3}T^{\rm charge}_{2}$. Typical charge qubit dephasing time has an order of ns~\cite{PhysRevLett.91.226804,PhysRevLett.95.090502}, such that the spin qubit dephasing time is likely to have an order of $\mu$s.

\section{Discussion and summary}

The longitudinal interaction mechanism between the spin qubit and photons in our paper is somewhat similar to that between the charge qubit and photons in a double quantum dot, where the two unequal interdot charge energies play an essential role~\cite{PhysRevLett.91.226804,PhysRevApplied.20.064005}, e.g., the double dot confining potential is asymmetrical.  Spin-photon interaction can also be mediated by a synthetic spin-orbit coupling~\cite{PhysRevX.12.021026,PhysRevApplied.18.064082,PhysRevB.86.035314,RL2019}, e.g., created by a gradient magnetic field.  The exchange gate~\cite{PhysRevX.12.021026} and iSWAP gate~\cite{PhysRevApplied.18.064082} between spin qubits can further be mediated via virtual microwave photons, and optimal operating regimes for an iSWAP gate were revealed~\cite{PhysRevApplied.18.064082}. Although our paper focuses on the static longitudinal spin-photon interaction, the dynamic longitudinal spin-photon interaction, i.e., induced by an external driving field, is also a very active research field~\cite{PhysRevApplied.20.064005,Bottcher:2022aa}. Let us also discuss on the magnitude of the transverse spin-photon interaction (in terms of length). The charge-photon interaction has an order of the quantum dot size, i.e., 40 nm in our paper. While the spin-photon interaction, mediated by the spin-orbit coupling, must be much smaller than the charge-photon interaction. Our calculated spin-photon interaction has an order of magnitude of 1 nm, hence is very reasonable.

In summary, via exactly solving the spin basis states, we obtain not only a transverse spin-photon interaction but also a longitudinal spin-photon interaction in an asymmetrical nanowire quantum dot. For both electron spin in an InSb nanowire quantum dot and hole pseudo spin in a Ge nanowire quantum dot, we calculate the dependences of the spin splitting, and both the transverse and the longitudinal spin-photon interactions on a series of external tunable parameters. For realistic spin-orbit coupling and realistic quantum dot parameters, the longitudinal interaction is at least one-order smaller than the transverse interaction. The longitudinal spin-photon interaction results from the confined effects of spin-orbit coupling and the asymmetrical confining potential. There is no longitudinal spin-photon interaction in a symmetrical quantum dot. The longitudinal spin-electric interaction demonstrated here may provide a mechanism for the charge noise induced spin dephasing in semiconductor quantum dot~\cite{RL2018c,RL2020}. 

\section*{Acknowledgements}
This work was supported by the National Natural Science Foundation of China Grant No.~11404020, the Project from the Department of Education of Hebei Province Grant No. QN2019057, and the Starting up Foundation from Yanshan University Grant No. BL18043.

\bibliography{Ref_Hole_spin}

\begin{thebibliography}{73}%
\makeatletter
\providecommand \@ifxundefined [1]{%
 \@ifx{#1\undefined}
}%
\providecommand \@ifnum [1]{%
 \ifnum #1\expandafter \@firstoftwo
 \else \expandafter \@secondoftwo
 \fi
}%
\providecommand \@ifx [1]{%
 \ifx #1\expandafter \@firstoftwo
 \else \expandafter \@secondoftwo
 \fi
}%
\providecommand \natexlab [1]{#1}%
\providecommand \enquote  [1]{``#1''}%
\providecommand \bibnamefont  [1]{#1}%
\providecommand \bibfnamefont [1]{#1}%
\providecommand \citenamefont [1]{#1}%
\providecommand \href@noop [0]{\@secondoftwo}%
\providecommand \href [0]{\begingroup \@sanitize@url \@href}%
\providecommand \@href[1]{\@@startlink{#1}\@@href}%
\providecommand \@@href[1]{\endgroup#1\@@endlink}%
\providecommand \@sanitize@url [0]{\catcode `\\12\catcode `\$12\catcode
  `\&12\catcode `\#12\catcode `\^12\catcode `\_12\catcode `\%12\relax}%
\providecommand \@@startlink[1]{}%
\providecommand \@@endlink[0]{}%
\providecommand \url  [0]{\begingroup\@sanitize@url \@url }%
\providecommand \@url [1]{\endgroup\@href {#1}{\urlprefix }}%
\providecommand \urlprefix  [0]{URL }%
\providecommand \Eprint [0]{\href }%
\providecommand \doibase [0]{https://doi.org/}%
\providecommand \selectlanguage [0]{\@gobble}%
\providecommand \bibinfo  [0]{\@secondoftwo}%
\providecommand \bibfield  [0]{\@secondoftwo}%
\providecommand \translation [1]{[#1]}%
\providecommand \BibitemOpen [0]{}%
\providecommand \bibitemStop [0]{}%
\providecommand \bibitemNoStop [0]{.\EOS\space}%
\providecommand \EOS [0]{\spacefactor3000\relax}%
\providecommand \BibitemShut  [1]{\csname bibitem#1\endcsname}%
\let\auto@bib@innerbib\@empty
\bibitem [{\citenamefont {Loss}\ and\ \citenamefont
  {DiVincenzo}(1998)}]{loss1998quantum}%
  \BibitemOpen
  \bibfield  {author} {\bibinfo {author} {\bibfnamefont {D.}~\bibnamefont
  {Loss}}\ and\ \bibinfo {author} {\bibfnamefont {D.~P.}\ \bibnamefont
  {DiVincenzo}},\ }\bibfield  {title} {\bibinfo {title} {Quantum computation
  with quantum dots},\ }\href {https://doi.org/10.1103/PhysRevA.57.120}
  {\bibfield  {journal} {\bibinfo  {journal} {Phys. Rev. A}\ }\textbf {\bibinfo
  {volume} {57}},\ \bibinfo {pages} {120} (\bibinfo {year} {1998})}\BibitemShut
  {NoStop}%
\bibitem [{\citenamefont {Vandersypen}\ and\ \citenamefont
  {Eriksson}(2019)}]{vandersypen2019}%
  \BibitemOpen
  \bibfield  {author} {\bibinfo {author} {\bibfnamefont {L.~M.~K.}\
  \bibnamefont {Vandersypen}}\ and\ \bibinfo {author} {\bibfnamefont {M.~A.}\
  \bibnamefont {Eriksson}},\ }\bibfield  {title} {\bibinfo {title} {Quantum
  computing with semiconductor spins},\ }\href
  {https://doi.org/10.1063/PT.3.4270} {\bibfield  {journal} {\bibinfo
  {journal} {Physics Today}\ }\textbf {\bibinfo {volume} {72}},\ \bibinfo
  {pages} {38} (\bibinfo {year} {2019})}\BibitemShut {NoStop}%
\bibitem [{\citenamefont {Scappucci}\ \emph {et~al.}(2021)\citenamefont
  {Scappucci}, \citenamefont {Kloeffel}, \citenamefont {Zwanenburg},
  \citenamefont {Loss}, \citenamefont {Myronov}, \citenamefont {Zhang},
  \citenamefont {De~Franceschi}, \citenamefont {Katsaros},\ and\ \citenamefont
  {Veldhorst}}]{Scappucci:2021vk}%
  \BibitemOpen
  \bibfield  {author} {\bibinfo {author} {\bibfnamefont {G.}~\bibnamefont
  {Scappucci}}, \bibinfo {author} {\bibfnamefont {C.}~\bibnamefont {Kloeffel}},
  \bibinfo {author} {\bibfnamefont {F.~A.}\ \bibnamefont {Zwanenburg}},
  \bibinfo {author} {\bibfnamefont {D.}~\bibnamefont {Loss}}, \bibinfo {author}
  {\bibfnamefont {M.}~\bibnamefont {Myronov}}, \bibinfo {author} {\bibfnamefont
  {J.-J.}\ \bibnamefont {Zhang}}, \bibinfo {author} {\bibfnamefont
  {S.}~\bibnamefont {De~Franceschi}}, \bibinfo {author} {\bibfnamefont
  {G.}~\bibnamefont {Katsaros}},\ and\ \bibinfo {author} {\bibfnamefont
  {M.}~\bibnamefont {Veldhorst}},\ }\bibfield  {title} {\bibinfo {title} {The
  germanium quantum information route},\ }\href
  {https://doi.org/10.1038/s41578-020-00262-z} {\bibfield  {journal} {\bibinfo
  {journal} {Nature Reviews Materials}\ }\textbf {\bibinfo {volume} {6}},\
  \bibinfo {pages} {926} (\bibinfo {year} {2021})}\BibitemShut {NoStop}%
\bibitem [{\citenamefont {Burkard}\ \emph {et~al.}(2023)\citenamefont
  {Burkard}, \citenamefont {Ladd}, \citenamefont {Pan}, \citenamefont
  {Nichol},\ and\ \citenamefont {Petta}}]{RevModPhys.95.025003}%
  \BibitemOpen
  \bibfield  {author} {\bibinfo {author} {\bibfnamefont {G.}~\bibnamefont
  {Burkard}}, \bibinfo {author} {\bibfnamefont {T.~D.}\ \bibnamefont {Ladd}},
  \bibinfo {author} {\bibfnamefont {A.}~\bibnamefont {Pan}}, \bibinfo {author}
  {\bibfnamefont {J.~M.}\ \bibnamefont {Nichol}},\ and\ \bibinfo {author}
  {\bibfnamefont {J.~R.}\ \bibnamefont {Petta}},\ }\bibfield  {title} {\bibinfo
  {title} {Semiconductor spin qubits},\ }\href
  {https://doi.org/10.1103/RevModPhys.95.025003} {\bibfield  {journal}
  {\bibinfo  {journal} {Rev. Mod. Phys.}\ }\textbf {\bibinfo {volume} {95}},\
  \bibinfo {pages} {025003} (\bibinfo {year} {2023})}\BibitemShut {NoStop}%
\bibitem [{\citenamefont {Rashba}\ and\ \citenamefont
  {Efros}(2003)}]{PhysRevLett91126405}%
  \BibitemOpen
  \bibfield  {author} {\bibinfo {author} {\bibfnamefont {E.~I.}\ \bibnamefont
  {Rashba}}\ and\ \bibinfo {author} {\bibfnamefont {A.~L.}\ \bibnamefont
  {Efros}},\ }\bibfield  {title} {\bibinfo {title} {Orbital mechanisms of
  electron-spin manipulation by an electric field},\ }\href
  {https://doi.org/10.1103/PhysRevLett.91.126405} {\bibfield  {journal}
  {\bibinfo  {journal} {Phys. Rev. Lett.}\ }\textbf {\bibinfo {volume} {91}},\
  \bibinfo {pages} {126405} (\bibinfo {year} {2003})}\BibitemShut {NoStop}%
\bibitem [{\citenamefont {Golovach}\ \emph {et~al.}(2006)\citenamefont
  {Golovach}, \citenamefont {Borhani},\ and\ \citenamefont
  {Loss}}]{golovach2006electric}%
  \BibitemOpen
  \bibfield  {author} {\bibinfo {author} {\bibfnamefont {V.~N.}\ \bibnamefont
  {Golovach}}, \bibinfo {author} {\bibfnamefont {M.}~\bibnamefont {Borhani}},\
  and\ \bibinfo {author} {\bibfnamefont {D.}~\bibnamefont {Loss}},\ }\bibfield
  {title} {\bibinfo {title} {Electric-dipole-induced spin resonance in quantum
  dots},\ }\href {https://doi.org/10.1103/PhysRevB.74.165319} {\bibfield
  {journal} {\bibinfo  {journal} {Phys. Rev. B}\ }\textbf {\bibinfo {volume}
  {74}},\ \bibinfo {pages} {165319} (\bibinfo {year} {2006})}\BibitemShut
  {NoStop}%
\bibitem [{\citenamefont {Bulaev}\ and\ \citenamefont
  {Loss}(2007)}]{PhysRevLett.98.097202}%
  \BibitemOpen
  \bibfield  {author} {\bibinfo {author} {\bibfnamefont {D.~V.}\ \bibnamefont
  {Bulaev}}\ and\ \bibinfo {author} {\bibfnamefont {D.}~\bibnamefont {Loss}},\
  }\bibfield  {title} {\bibinfo {title} {Electric dipole spin resonance for
  heavy holes in quantum dots},\ }\href
  {https://doi.org/10.1103/PhysRevLett.98.097202} {\bibfield  {journal}
  {\bibinfo  {journal} {Phys. Rev. Lett.}\ }\textbf {\bibinfo {volume} {98}},\
  \bibinfo {pages} {097202} (\bibinfo {year} {2007})}\BibitemShut {NoStop}%
\bibitem [{\citenamefont {Li}\ \emph {et~al.}(2013)\citenamefont {Li},
  \citenamefont {You}, \citenamefont {Sun},\ and\ \citenamefont
  {Nori}}]{RL2013}%
  \BibitemOpen
  \bibfield  {author} {\bibinfo {author} {\bibfnamefont {R.}~\bibnamefont
  {Li}}, \bibinfo {author} {\bibfnamefont {J.~Q.}\ \bibnamefont {You}},
  \bibinfo {author} {\bibfnamefont {C.~P.}\ \bibnamefont {Sun}},\ and\ \bibinfo
  {author} {\bibfnamefont {F.}~\bibnamefont {Nori}},\ }\bibfield  {title}
  {\bibinfo {title} {Controlling a nanowire spin-orbit qubit via
  electric-dipole spin resonance},\ }\href
  {https://doi.org/10.1103/PhysRevLett.111.086805} {\bibfield  {journal}
  {\bibinfo  {journal} {Phys. Rev. Lett.}\ }\textbf {\bibinfo {volume} {111}},\
  \bibinfo {pages} {086805} (\bibinfo {year} {2013})}\BibitemShut {NoStop}%
\bibitem [{\citenamefont {Romh\'anyi}\ \emph {et~al.}(2015)\citenamefont
  {Romh\'anyi}, \citenamefont {Burkard},\ and\ \citenamefont
  {P\'alyi}}]{PhysRevB.92.054422}%
  \BibitemOpen
  \bibfield  {author} {\bibinfo {author} {\bibfnamefont {J.}~\bibnamefont
  {Romh\'anyi}}, \bibinfo {author} {\bibfnamefont {G.}~\bibnamefont
  {Burkard}},\ and\ \bibinfo {author} {\bibfnamefont {A.}~\bibnamefont
  {P\'alyi}},\ }\bibfield  {title} {\bibinfo {title} {Subharmonic transitions
  and bloch-siegert shift in electrically driven spin resonance},\ }\href
  {https://doi.org/10.1103/PhysRevB.92.054422} {\bibfield  {journal} {\bibinfo
  {journal} {Phys. Rev. B}\ }\textbf {\bibinfo {volume} {92}},\ \bibinfo
  {pages} {054422} (\bibinfo {year} {2015})}\BibitemShut {NoStop}%
\bibitem [{\citenamefont {Wang}\ \emph {et~al.}(2021)\citenamefont {Wang},
  \citenamefont {Marcellina}, \citenamefont {Hamilton}, \citenamefont {Cullen},
  \citenamefont {Rogge}, \citenamefont {Salfi},\ and\ \citenamefont
  {Culcer}}]{Wang:2021wc}%
  \BibitemOpen
  \bibfield  {author} {\bibinfo {author} {\bibfnamefont {Z.}~\bibnamefont
  {Wang}}, \bibinfo {author} {\bibfnamefont {E.}~\bibnamefont {Marcellina}},
  \bibinfo {author} {\bibfnamefont {A.~R.}\ \bibnamefont {Hamilton}}, \bibinfo
  {author} {\bibfnamefont {J.~H.}\ \bibnamefont {Cullen}}, \bibinfo {author}
  {\bibfnamefont {S.}~\bibnamefont {Rogge}}, \bibinfo {author} {\bibfnamefont
  {J.}~\bibnamefont {Salfi}},\ and\ \bibinfo {author} {\bibfnamefont
  {D.}~\bibnamefont {Culcer}},\ }\bibfield  {title} {\bibinfo {title} {Optimal
  operation points for ultrafast, highly coherent {Ge} hole spin-orbit
  qubits},\ }\href {https://doi.org/10.1038/s41534-021-00386-2} {\bibfield
  {journal} {\bibinfo  {journal} {npj Quantum Information}\ }\textbf {\bibinfo
  {volume} {7}},\ \bibinfo {pages} {54} (\bibinfo {year} {2021})}\BibitemShut
  {NoStop}%
\bibitem [{\citenamefont {Khomitsky}\ \emph {et~al.}(2020)\citenamefont
  {Khomitsky}, \citenamefont {Lavrukhina},\ and\ \citenamefont
  {Sherman}}]{PhysRevApplied.14.014090}%
  \BibitemOpen
  \bibfield  {author} {\bibinfo {author} {\bibfnamefont {D.}~\bibnamefont
  {Khomitsky}}, \bibinfo {author} {\bibfnamefont {E.}~\bibnamefont
  {Lavrukhina}},\ and\ \bibinfo {author} {\bibfnamefont {E.}~\bibnamefont
  {Sherman}},\ }\bibfield  {title} {\bibinfo {title} {Spin rotation by resonant
  electric field in few-level quantum dots: Floquet dynamics and tunneling},\
  }\href {https://doi.org/10.1103/PhysRevApplied.14.014090} {\bibfield
  {journal} {\bibinfo  {journal} {Phys. Rev. Applied}\ }\textbf {\bibinfo
  {volume} {14}},\ \bibinfo {pages} {014090} (\bibinfo {year}
  {2020})}\BibitemShut {NoStop}%
\bibitem [{\citenamefont {Burkard}\ \emph {et~al.}(1999)\citenamefont
  {Burkard}, \citenamefont {Loss},\ and\ \citenamefont
  {DiVincenzo}}]{burkard1999coupled}%
  \BibitemOpen
  \bibfield  {author} {\bibinfo {author} {\bibfnamefont {G.}~\bibnamefont
  {Burkard}}, \bibinfo {author} {\bibfnamefont {D.}~\bibnamefont {Loss}},\ and\
  \bibinfo {author} {\bibfnamefont {D.~P.}\ \bibnamefont {DiVincenzo}},\
  }\bibfield  {title} {\bibinfo {title} {Coupled quantum dots as quantum
  gates},\ }\href {https://doi.org/10.1103/PhysRevB.59.2070} {\bibfield
  {journal} {\bibinfo  {journal} {Phys. Rev. B}\ }\textbf {\bibinfo {volume}
  {59}},\ \bibinfo {pages} {2070} (\bibinfo {year} {1999})}\BibitemShut
  {NoStop}%
\bibitem [{\citenamefont {Hu}\ and\ \citenamefont
  {Das~Sarma}(2000)}]{hu2000hilbert}%
  \BibitemOpen
  \bibfield  {author} {\bibinfo {author} {\bibfnamefont {X.}~\bibnamefont
  {Hu}}\ and\ \bibinfo {author} {\bibfnamefont {S.}~\bibnamefont {Das~Sarma}},\
  }\bibfield  {title} {\bibinfo {title} {Hilbert-space structure of a
  solid-state quantum computer: Two-electron states of a double-quantum-dot
  artificial molecule},\ }\href {https://doi.org/10.1103/PhysRevA.61.062301}
  {\bibfield  {journal} {\bibinfo  {journal} {Phys. Rev. A}\ }\textbf {\bibinfo
  {volume} {61}},\ \bibinfo {pages} {062301} (\bibinfo {year}
  {2000})}\BibitemShut {NoStop}%
\bibitem [{\citenamefont {Kawakami}\ \emph {et~al.}(2016)\citenamefont
  {Kawakami}, \citenamefont {Jullien}, \citenamefont {Scarlino}, \citenamefont
  {Ward}, \citenamefont {Savage}, \citenamefont {Lagally}, \citenamefont
  {Dobrovitski}, \citenamefont {Friesen}, \citenamefont {Coppersmith},
  \citenamefont {Eriksson},\ and\ \citenamefont
  {Vandersypen}}]{doi:10.1073/pnas.1603251113}%
  \BibitemOpen
  \bibfield  {author} {\bibinfo {author} {\bibfnamefont {E.}~\bibnamefont
  {Kawakami}}, \bibinfo {author} {\bibfnamefont {T.}~\bibnamefont {Jullien}},
  \bibinfo {author} {\bibfnamefont {P.}~\bibnamefont {Scarlino}}, \bibinfo
  {author} {\bibfnamefont {D.~R.}\ \bibnamefont {Ward}}, \bibinfo {author}
  {\bibfnamefont {D.~E.}\ \bibnamefont {Savage}}, \bibinfo {author}
  {\bibfnamefont {M.~G.}\ \bibnamefont {Lagally}}, \bibinfo {author}
  {\bibfnamefont {V.~V.}\ \bibnamefont {Dobrovitski}}, \bibinfo {author}
  {\bibfnamefont {M.}~\bibnamefont {Friesen}}, \bibinfo {author} {\bibfnamefont
  {S.~N.}\ \bibnamefont {Coppersmith}}, \bibinfo {author} {\bibfnamefont
  {M.~A.}\ \bibnamefont {Eriksson}},\ and\ \bibinfo {author} {\bibfnamefont
  {L.~M.~K.}\ \bibnamefont {Vandersypen}},\ }\bibfield  {title} {\bibinfo
  {title} {Gate fidelity and coherence of an electron spin in an {Si/SiGe}
  quantum dot with micromagnet},\ }\href
  {https://doi.org/10.1073/pnas.1603251113} {\bibfield  {journal} {\bibinfo
  {journal} {Proceedings of the National Academy of Sciences}\ }\textbf
  {\bibinfo {volume} {113}},\ \bibinfo {pages} {11738} (\bibinfo {year}
  {2016})}\BibitemShut {NoStop}%
\bibitem [{\citenamefont {Yoneda}\ \emph {et~al.}(2018)\citenamefont {Yoneda},
  \citenamefont {Takeda}, \citenamefont {Otsuka}, \citenamefont {Nakajima},
  \citenamefont {Delbecq}, \citenamefont {Allison}, \citenamefont {Honda},
  \citenamefont {Kodera}, \citenamefont {Oda}, \citenamefont {Hoshi} \emph
  {et~al.}}]{yoneda2018}%
  \BibitemOpen
  \bibfield  {author} {\bibinfo {author} {\bibfnamefont {J.}~\bibnamefont
  {Yoneda}}, \bibinfo {author} {\bibfnamefont {K.}~\bibnamefont {Takeda}},
  \bibinfo {author} {\bibfnamefont {T.}~\bibnamefont {Otsuka}}, \bibinfo
  {author} {\bibfnamefont {T.}~\bibnamefont {Nakajima}}, \bibinfo {author}
  {\bibfnamefont {M.~R.}\ \bibnamefont {Delbecq}}, \bibinfo {author}
  {\bibfnamefont {G.}~\bibnamefont {Allison}}, \bibinfo {author} {\bibfnamefont
  {T.}~\bibnamefont {Honda}}, \bibinfo {author} {\bibfnamefont
  {T.}~\bibnamefont {Kodera}}, \bibinfo {author} {\bibfnamefont
  {S.}~\bibnamefont {Oda}}, \bibinfo {author} {\bibfnamefont {Y.}~\bibnamefont
  {Hoshi}}, \emph {et~al.},\ }\bibfield  {title} {\bibinfo {title} {A
  quantum-dot spin qubit with coherence limited by charge noise and fidelity
  higher than 99.9\%},\ }\href {https://doi.org/10.1038/s41565-017-0014-x}
  {\bibfield  {journal} {\bibinfo  {journal} {Nature nanotechnology}\ }\textbf
  {\bibinfo {volume} {13}},\ \bibinfo {pages} {102} (\bibinfo {year}
  {2018})}\BibitemShut {NoStop}%
\bibitem [{\citenamefont {Witzel}\ and\ \citenamefont
  {Das~Sarma}(2006)}]{witzel2006quantum}%
  \BibitemOpen
  \bibfield  {author} {\bibinfo {author} {\bibfnamefont {W.~M.}\ \bibnamefont
  {Witzel}}\ and\ \bibinfo {author} {\bibfnamefont {S.}~\bibnamefont
  {Das~Sarma}},\ }\bibfield  {title} {\bibinfo {title} {Quantum theory for
  electron spin decoherence induced by nuclear spin dynamics in semiconductor
  quantum computer architectures: Spectral diffusion of localized electron
  spins in the nuclear solid-state environment},\ }\href
  {https://doi.org/10.1103/PhysRevB.74.035322} {\bibfield  {journal} {\bibinfo
  {journal} {Phys. Rev. B}\ }\textbf {\bibinfo {volume} {74}},\ \bibinfo
  {pages} {035322} (\bibinfo {year} {2006})}\BibitemShut {NoStop}%
\bibitem [{\citenamefont {Yao}\ \emph {et~al.}(2006)\citenamefont {Yao},
  \citenamefont {Liu},\ and\ \citenamefont {Sham}}]{yao2006theory}%
  \BibitemOpen
  \bibfield  {author} {\bibinfo {author} {\bibfnamefont {W.}~\bibnamefont
  {Yao}}, \bibinfo {author} {\bibfnamefont {R.-B.}\ \bibnamefont {Liu}},\ and\
  \bibinfo {author} {\bibfnamefont {L.~J.}\ \bibnamefont {Sham}},\ }\bibfield
  {title} {\bibinfo {title} {Theory of electron spin decoherence by interacting
  nuclear spins in a quantum dot},\ }\href
  {https://doi.org/10.1103/PhysRevB.74.195301} {\bibfield  {journal} {\bibinfo
  {journal} {Phys. Rev. B}\ }\textbf {\bibinfo {volume} {74}},\ \bibinfo
  {pages} {195301} (\bibinfo {year} {2006})}\BibitemShut {NoStop}%
\bibitem [{\citenamefont {Khaetskii}\ and\ \citenamefont
  {Nazarov}(2000)}]{PhysRevB.61.12639}%
  \BibitemOpen
  \bibfield  {author} {\bibinfo {author} {\bibfnamefont {A.~V.}\ \bibnamefont
  {Khaetskii}}\ and\ \bibinfo {author} {\bibfnamefont {Y.~V.}\ \bibnamefont
  {Nazarov}},\ }\bibfield  {title} {\bibinfo {title} {Spin relaxation in
  semiconductor quantum dots},\ }\href
  {https://doi.org/10.1103/PhysRevB.61.12639} {\bibfield  {journal} {\bibinfo
  {journal} {Phys. Rev. B}\ }\textbf {\bibinfo {volume} {61}},\ \bibinfo
  {pages} {12639} (\bibinfo {year} {2000})}\BibitemShut {NoStop}%
\bibitem [{\citenamefont {Khaetskii}\ and\ \citenamefont
  {Nazarov}(2001)}]{PhysRevB.64.125316}%
  \BibitemOpen
  \bibfield  {author} {\bibinfo {author} {\bibfnamefont {A.~V.}\ \bibnamefont
  {Khaetskii}}\ and\ \bibinfo {author} {\bibfnamefont {Y.~V.}\ \bibnamefont
  {Nazarov}},\ }\bibfield  {title} {\bibinfo {title} {Spin-flip transitions
  between zeeman sublevels in semiconductor quantum dots},\ }\href
  {https://doi.org/10.1103/PhysRevB.64.125316} {\bibfield  {journal} {\bibinfo
  {journal} {Phys. Rev. B}\ }\textbf {\bibinfo {volume} {64}},\ \bibinfo
  {pages} {125316} (\bibinfo {year} {2001})}\BibitemShut {NoStop}%
\bibitem [{\citenamefont {Bermeister}\ \emph {et~al.}(2014)\citenamefont
  {Bermeister}, \citenamefont {Keith},\ and\ \citenamefont
  {Culcer}}]{10.1063/1.4901162}%
  \BibitemOpen
  \bibfield  {author} {\bibinfo {author} {\bibfnamefont {A.}~\bibnamefont
  {Bermeister}}, \bibinfo {author} {\bibfnamefont {D.}~\bibnamefont {Keith}},\
  and\ \bibinfo {author} {\bibfnamefont {D.}~\bibnamefont {Culcer}},\
  }\bibfield  {title} {\bibinfo {title} {{Charge noise, spin-orbit coupling,
  and dephasing of single-spin qubits}},\ }\href
  {https://doi.org/10.1063/1.4901162} {\bibfield  {journal} {\bibinfo
  {journal} {Applied Physics Letters}\ }\textbf {\bibinfo {volume} {105}},\
  \bibinfo {pages} {192102} (\bibinfo {year} {2014})}\BibitemShut {NoStop}%
\bibitem [{\citenamefont {Li}(2018)}]{RL2018c}%
  \BibitemOpen
  \bibfield  {author} {\bibinfo {author} {\bibfnamefont {R.}~\bibnamefont
  {Li}},\ }\bibfield  {title} {\bibinfo {title} {A spin dephasing mechanism
  mediated by the interplay between the spin-orbit coupling and the
  asymmetrical confining potential in a semiconductor quantum dot},\ }\href
  {https://doi.org/10.1088/1361-648X/aadcb8} {\bibfield  {journal} {\bibinfo
  {journal} {Journal of Physics: Condensed Matter}\ }\textbf {\bibinfo {volume}
  {30}},\ \bibinfo {pages} {395304} (\bibinfo {year} {2018})}\BibitemShut
  {NoStop}%
\bibitem [{\citenamefont {Li}(2020)}]{RL2020}%
  \BibitemOpen
  \bibfield  {author} {\bibinfo {author} {\bibfnamefont {R.}~\bibnamefont
  {Li}},\ }\bibfield  {title} {\bibinfo {title} {Charge noise induced spin
  dephasing in a nanowire double quantum dot with spin{\textendash}orbit
  coupling},\ }\href {https://doi.org/10.1088/1361-648x/ab4933} {\bibfield
  {journal} {\bibinfo  {journal} {Journal of Physics: Condensed Matter}\
  }\textbf {\bibinfo {volume} {32}},\ \bibinfo {pages} {025305} (\bibinfo
  {year} {2020})}\BibitemShut {NoStop}%
\bibitem [{\citenamefont {Scully}\ and\ \citenamefont
  {Zubairy}(1997)}]{scully1999quantum}%
  \BibitemOpen
  \bibfield  {author} {\bibinfo {author} {\bibfnamefont {M.~O.}\ \bibnamefont
  {Scully}}\ and\ \bibinfo {author} {\bibfnamefont {M.~S.}\ \bibnamefont
  {Zubairy}},\ }\href@noop {} {\emph {\bibinfo {title} {Quantum optics}}}\
  (\bibinfo  {publisher} {Cambridge University Press, Cambridge, England},\
  \bibinfo {year} {1997})\BibitemShut {NoStop}%
\bibitem [{\citenamefont {Jaynes}\ and\ \citenamefont
  {Cummings}(1963)}]{1443594}%
  \BibitemOpen
  \bibfield  {author} {\bibinfo {author} {\bibfnamefont {E.}~\bibnamefont
  {Jaynes}}\ and\ \bibinfo {author} {\bibfnamefont {F.}~\bibnamefont
  {Cummings}},\ }\bibfield  {title} {\bibinfo {title} {Comparison of quantum
  and semiclassical radiation theories with application to the beam maser},\
  }\href {https://doi.org/10.1109/PROC.1963.1664} {\bibfield  {journal}
  {\bibinfo  {journal} {Proceedings of the IEEE}\ }\textbf {\bibinfo {volume}
  {51}},\ \bibinfo {pages} {89} (\bibinfo {year} {1963})}\BibitemShut {NoStop}%
\bibitem [{\citenamefont {Rabi}(1936)}]{PhysRev.49.324}%
  \BibitemOpen
  \bibfield  {author} {\bibinfo {author} {\bibfnamefont {I.~I.}\ \bibnamefont
  {Rabi}},\ }\bibfield  {title} {\bibinfo {title} {On the process of space
  quantization},\ }\href {https://doi.org/10.1103/PhysRev.49.324} {\bibfield
  {journal} {\bibinfo  {journal} {Phys. Rev.}\ }\textbf {\bibinfo {volume}
  {49}},\ \bibinfo {pages} {324} (\bibinfo {year} {1936})}\BibitemShut
  {NoStop}%
\bibitem [{\citenamefont {Trif}\ \emph {et~al.}(2008)\citenamefont {Trif},
  \citenamefont {Golovach},\ and\ \citenamefont {Loss}}]{trif2008spin}%
  \BibitemOpen
  \bibfield  {author} {\bibinfo {author} {\bibfnamefont {M.}~\bibnamefont
  {Trif}}, \bibinfo {author} {\bibfnamefont {V.~N.}\ \bibnamefont {Golovach}},\
  and\ \bibinfo {author} {\bibfnamefont {D.}~\bibnamefont {Loss}},\ }\bibfield
  {title} {\bibinfo {title} {{Spin dynamics in InAs nanowire quantum dots
  coupled to a transmission line}},\ }\href
  {https://doi.org/10.1103/PhysRevB.77.045434} {\bibfield  {journal} {\bibinfo
  {journal} {Phys. Rev. B}\ }\textbf {\bibinfo {volume} {77}},\ \bibinfo
  {pages} {045434} (\bibinfo {year} {2008})}\BibitemShut {NoStop}%
\bibitem [{\citenamefont {Hu}\ \emph {et~al.}(2012)\citenamefont {Hu},
  \citenamefont {Liu},\ and\ \citenamefont {Nori}}]{PhysRevB.86.035314}%
  \BibitemOpen
  \bibfield  {author} {\bibinfo {author} {\bibfnamefont {X.}~\bibnamefont
  {Hu}}, \bibinfo {author} {\bibfnamefont {Y.-x.}\ \bibnamefont {Liu}},\ and\
  \bibinfo {author} {\bibfnamefont {F.}~\bibnamefont {Nori}},\ }\bibfield
  {title} {\bibinfo {title} {Strong coupling of a spin qubit to a
  superconducting stripline cavity},\ }\href
  {https://doi.org/10.1103/PhysRevB.86.035314} {\bibfield  {journal} {\bibinfo
  {journal} {Phys. Rev. B}\ }\textbf {\bibinfo {volume} {86}},\ \bibinfo
  {pages} {035314} (\bibinfo {year} {2012})}\BibitemShut {NoStop}%
\bibitem [{\citenamefont {Kloeffel}\ \emph {et~al.}(2013)\citenamefont
  {Kloeffel}, \citenamefont {Trif}, \citenamefont {Stano},\ and\ \citenamefont
  {Loss}}]{PhysRevB.88.241405}%
  \BibitemOpen
  \bibfield  {author} {\bibinfo {author} {\bibfnamefont {C.}~\bibnamefont
  {Kloeffel}}, \bibinfo {author} {\bibfnamefont {M.}~\bibnamefont {Trif}},
  \bibinfo {author} {\bibfnamefont {P.}~\bibnamefont {Stano}},\ and\ \bibinfo
  {author} {\bibfnamefont {D.}~\bibnamefont {Loss}},\ }\bibfield  {title}
  {\bibinfo {title} {Circuit {QED} with hole-spin qubits in {G}e/{S}i nanowire
  quantum dots},\ }\href {https://doi.org/10.1103/PhysRevB.88.241405}
  {\bibfield  {journal} {\bibinfo  {journal} {Phys. Rev. B}\ }\textbf {\bibinfo
  {volume} {88}},\ \bibinfo {pages} {241405} (\bibinfo {year}
  {2013})}\BibitemShut {NoStop}%
\bibitem [{\citenamefont {Mutter}\ and\ \citenamefont
  {Burkard}(2020)}]{PhysRevB.102.205412}%
  \BibitemOpen
  \bibfield  {author} {\bibinfo {author} {\bibfnamefont {P.~M.}\ \bibnamefont
  {Mutter}}\ and\ \bibinfo {author} {\bibfnamefont {G.}~\bibnamefont
  {Burkard}},\ }\bibfield  {title} {\bibinfo {title} {Cavity control over
  heavy-hole spin qubits in inversion-symmetric crystals},\ }\href
  {https://doi.org/10.1103/PhysRevB.102.205412} {\bibfield  {journal} {\bibinfo
   {journal} {Phys. Rev. B}\ }\textbf {\bibinfo {volume} {102}},\ \bibinfo
  {pages} {205412} (\bibinfo {year} {2020})}\BibitemShut {NoStop}%
\bibitem [{\citenamefont {Jin}\ \emph {et~al.}(2012)\citenamefont {Jin},
  \citenamefont {Marthaler}, \citenamefont {Shnirman},\ and\ \citenamefont
  {Sch\"on}}]{PhysRevLett.108.190506}%
  \BibitemOpen
  \bibfield  {author} {\bibinfo {author} {\bibfnamefont {P.-Q.}\ \bibnamefont
  {Jin}}, \bibinfo {author} {\bibfnamefont {M.}~\bibnamefont {Marthaler}},
  \bibinfo {author} {\bibfnamefont {A.}~\bibnamefont {Shnirman}},\ and\
  \bibinfo {author} {\bibfnamefont {G.}~\bibnamefont {Sch\"on}},\ }\bibfield
  {title} {\bibinfo {title} {Strong coupling of spin qubits to a transmission
  line resonator},\ }\href {https://doi.org/10.1103/PhysRevLett.108.190506}
  {\bibfield  {journal} {\bibinfo  {journal} {Phys. Rev. Lett.}\ }\textbf
  {\bibinfo {volume} {108}},\ \bibinfo {pages} {190506} (\bibinfo {year}
  {2012})}\BibitemShut {NoStop}%
\bibitem [{\citenamefont {Li}\ \emph {et~al.}(2018{\natexlab{a}})\citenamefont
  {Li}, \citenamefont {Li}, \citenamefont {Gao}, \citenamefont {Li},
  \citenamefont {Xu}, \citenamefont {Wang}, \citenamefont {Liu}, \citenamefont
  {Cao}, \citenamefont {Xiao}, \citenamefont {Wang}, \citenamefont {Zhang},
  \citenamefont {Guo},\ and\ \citenamefont {Guo}}]{Li:2018ab}%
  \BibitemOpen
  \bibfield  {author} {\bibinfo {author} {\bibfnamefont {Y.}~\bibnamefont
  {Li}}, \bibinfo {author} {\bibfnamefont {S.-X.}\ \bibnamefont {Li}}, \bibinfo
  {author} {\bibfnamefont {F.}~\bibnamefont {Gao}}, \bibinfo {author}
  {\bibfnamefont {H.-O.}\ \bibnamefont {Li}}, \bibinfo {author} {\bibfnamefont
  {G.}~\bibnamefont {Xu}}, \bibinfo {author} {\bibfnamefont {K.}~\bibnamefont
  {Wang}}, \bibinfo {author} {\bibfnamefont {D.}~\bibnamefont {Liu}}, \bibinfo
  {author} {\bibfnamefont {G.}~\bibnamefont {Cao}}, \bibinfo {author}
  {\bibfnamefont {M.}~\bibnamefont {Xiao}}, \bibinfo {author} {\bibfnamefont
  {T.}~\bibnamefont {Wang}}, \bibinfo {author} {\bibfnamefont {J.-J.}\
  \bibnamefont {Zhang}}, \bibinfo {author} {\bibfnamefont {G.-C.}\ \bibnamefont
  {Guo}},\ and\ \bibinfo {author} {\bibfnamefont {G.-P.}\ \bibnamefont {Guo}},\
  }\bibfield  {title} {\bibinfo {title} {Coupling a germanium hut wire hole
  quantum dot to a superconducting microwave resonator},\ }\href
  {https://doi.org/10.1021/acs.nanolett.8b00272} {\bibfield  {journal}
  {\bibinfo  {journal} {Nano Letters}\ }\textbf {\bibinfo {volume} {18}},\
  \bibinfo {pages} {2091} (\bibinfo {year} {2018}{\natexlab{a}})}\BibitemShut
  {NoStop}%
\bibitem [{\citenamefont {Burkard}\ \emph {et~al.}(2020)\citenamefont
  {Burkard}, \citenamefont {Gullans}, \citenamefont {Mi},\ and\ \citenamefont
  {Petta}}]{Burkard:2020aa}%
  \BibitemOpen
  \bibfield  {author} {\bibinfo {author} {\bibfnamefont {G.}~\bibnamefont
  {Burkard}}, \bibinfo {author} {\bibfnamefont {M.~J.}\ \bibnamefont
  {Gullans}}, \bibinfo {author} {\bibfnamefont {X.}~\bibnamefont {Mi}},\ and\
  \bibinfo {author} {\bibfnamefont {J.~R.}\ \bibnamefont {Petta}},\ }\bibfield
  {title} {\bibinfo {title} {Superconductor--semiconductor hybrid-circuit
  quantum electrodynamics},\ }\href {https://doi.org/10.1038/s42254-019-0135-2}
  {\bibfield  {journal} {\bibinfo  {journal} {Nature Reviews Physics}\ }\textbf
  {\bibinfo {volume} {2}},\ \bibinfo {pages} {129} (\bibinfo {year}
  {2020})}\BibitemShut {NoStop}%
\bibitem [{\citenamefont {Michal}\ \emph {et~al.}(2023)\citenamefont {Michal},
  \citenamefont {Abadillo-Uriel}, \citenamefont {Zihlmann}, \citenamefont
  {Maurand}, \citenamefont {Niquet},\ and\ \citenamefont
  {Filippone}}]{PhysRevB.107.L041303}%
  \BibitemOpen
  \bibfield  {author} {\bibinfo {author} {\bibfnamefont {V.~P.}\ \bibnamefont
  {Michal}}, \bibinfo {author} {\bibfnamefont {J.~C.}\ \bibnamefont
  {Abadillo-Uriel}}, \bibinfo {author} {\bibfnamefont {S.}~\bibnamefont
  {Zihlmann}}, \bibinfo {author} {\bibfnamefont {R.}~\bibnamefont {Maurand}},
  \bibinfo {author} {\bibfnamefont {Y.-M.}\ \bibnamefont {Niquet}},\ and\
  \bibinfo {author} {\bibfnamefont {M.}~\bibnamefont {Filippone}},\ }\bibfield
  {title} {\bibinfo {title} {Tunable hole spin-photon interaction based on
  $\mathtt{g}$-matrix modulation},\ }\href
  {https://doi.org/10.1103/PhysRevB.107.L041303} {\bibfield  {journal}
  {\bibinfo  {journal} {Phys. Rev. B}\ }\textbf {\bibinfo {volume} {107}},\
  \bibinfo {pages} {L041303} (\bibinfo {year} {2023})}\BibitemShut {NoStop}%
\bibitem [{\citenamefont {Bosco}\ \emph {et~al.}(2022)\citenamefont {Bosco},
  \citenamefont {Scarlino}, \citenamefont {Klinovaja},\ and\ \citenamefont
  {Loss}}]{PhysRevLett.129.066801}%
  \BibitemOpen
  \bibfield  {author} {\bibinfo {author} {\bibfnamefont {S.}~\bibnamefont
  {Bosco}}, \bibinfo {author} {\bibfnamefont {P.}~\bibnamefont {Scarlino}},
  \bibinfo {author} {\bibfnamefont {J.}~\bibnamefont {Klinovaja}},\ and\
  \bibinfo {author} {\bibfnamefont {D.}~\bibnamefont {Loss}},\ }\bibfield
  {title} {\bibinfo {title} {Fully tunable longitudinal spin-photon
  interactions in {Si} and {Ge} quantum dots},\ }\href
  {https://doi.org/10.1103/PhysRevLett.129.066801} {\bibfield  {journal}
  {\bibinfo  {journal} {Phys. Rev. Lett.}\ }\textbf {\bibinfo {volume} {129}},\
  \bibinfo {pages} {066801} (\bibinfo {year} {2022})}\BibitemShut {NoStop}%
\bibitem [{\citenamefont {Harvey-Collard}\ \emph {et~al.}(2022)\citenamefont
  {Harvey-Collard}, \citenamefont {Dijkema}, \citenamefont {Zheng},
  \citenamefont {Sammak}, \citenamefont {Scappucci},\ and\ \citenamefont
  {Vandersypen}}]{PhysRevX.12.021026}%
  \BibitemOpen
  \bibfield  {author} {\bibinfo {author} {\bibfnamefont {P.}~\bibnamefont
  {Harvey-Collard}}, \bibinfo {author} {\bibfnamefont {J.}~\bibnamefont
  {Dijkema}}, \bibinfo {author} {\bibfnamefont {G.}~\bibnamefont {Zheng}},
  \bibinfo {author} {\bibfnamefont {A.}~\bibnamefont {Sammak}}, \bibinfo
  {author} {\bibfnamefont {G.}~\bibnamefont {Scappucci}},\ and\ \bibinfo
  {author} {\bibfnamefont {L.~M.~K.}\ \bibnamefont {Vandersypen}},\ }\bibfield
  {title} {\bibinfo {title} {Coherent spin-spin coupling mediated by virtual
  microwave photons},\ }\href {https://doi.org/10.1103/PhysRevX.12.021026}
  {\bibfield  {journal} {\bibinfo  {journal} {Phys. Rev. X}\ }\textbf {\bibinfo
  {volume} {12}},\ \bibinfo {pages} {021026} (\bibinfo {year}
  {2022})}\BibitemShut {NoStop}%
\bibitem [{\citenamefont {Young}\ \emph {et~al.}(2022)\citenamefont {Young},
  \citenamefont {Jacobson},\ and\ \citenamefont
  {Petta}}]{PhysRevApplied.18.064082}%
  \BibitemOpen
  \bibfield  {author} {\bibinfo {author} {\bibfnamefont {S.~M.}\ \bibnamefont
  {Young}}, \bibinfo {author} {\bibfnamefont {N.~T.}\ \bibnamefont
  {Jacobson}},\ and\ \bibinfo {author} {\bibfnamefont {J.~R.}\ \bibnamefont
  {Petta}},\ }\bibfield  {title} {\bibinfo {title} {Optimal control of a
  cavity-mediated $\mathrm{i}\mathrm{SWAP}$ gate between silicon spin qubits},\
  }\href {https://doi.org/10.1103/PhysRevApplied.18.064082} {\bibfield
  {journal} {\bibinfo  {journal} {Phys. Rev. Appl.}\ }\textbf {\bibinfo
  {volume} {18}},\ \bibinfo {pages} {064082} (\bibinfo {year}
  {2022})}\BibitemShut {NoStop}%
\bibitem [{\citenamefont {Imamog\ifmmode\bar\else\textasciimacron\fi{}lu}\
  \emph {et~al.}(1999)\citenamefont
  {Imamog\ifmmode\bar\else\textasciimacron\fi{}lu}, \citenamefont {Awschalom},
  \citenamefont {Burkard}, \citenamefont {DiVincenzo}, \citenamefont {Loss},
  \citenamefont {Sherwin},\ and\ \citenamefont {Small}}]{PhysRevLett.83.4204}%
  \BibitemOpen
  \bibfield  {author} {\bibinfo {author} {\bibfnamefont {A.}~\bibnamefont
  {Imamog\ifmmode\bar\else\textasciimacron\fi{}lu}}, \bibinfo {author}
  {\bibfnamefont {D.~D.}\ \bibnamefont {Awschalom}}, \bibinfo {author}
  {\bibfnamefont {G.}~\bibnamefont {Burkard}}, \bibinfo {author} {\bibfnamefont
  {D.~P.}\ \bibnamefont {DiVincenzo}}, \bibinfo {author} {\bibfnamefont
  {D.}~\bibnamefont {Loss}}, \bibinfo {author} {\bibfnamefont {M.}~\bibnamefont
  {Sherwin}},\ and\ \bibinfo {author} {\bibfnamefont {A.}~\bibnamefont
  {Small}},\ }\bibfield  {title} {\bibinfo {title} {Quantum information
  processing using quantum dot spins and cavity qed},\ }\href
  {https://doi.org/10.1103/PhysRevLett.83.4204} {\bibfield  {journal} {\bibinfo
   {journal} {Phys. Rev. Lett.}\ }\textbf {\bibinfo {volume} {83}},\ \bibinfo
  {pages} {4204} (\bibinfo {year} {1999})}\BibitemShut {NoStop}%
\bibitem [{\citenamefont {Zheng}\ and\ \citenamefont
  {Guo}(2000)}]{PhysRevLett.85.2392}%
  \BibitemOpen
  \bibfield  {author} {\bibinfo {author} {\bibfnamefont {S.-B.}\ \bibnamefont
  {Zheng}}\ and\ \bibinfo {author} {\bibfnamefont {G.-C.}\ \bibnamefont
  {Guo}},\ }\bibfield  {title} {\bibinfo {title} {{Efficient Scheme for
  Two-Atom Entanglement and Quantum Information Processing in Cavity QED}},\
  }\href {https://doi.org/10.1103/PhysRevLett.85.2392} {\bibfield  {journal}
  {\bibinfo  {journal} {Phys. Rev. Lett.}\ }\textbf {\bibinfo {volume} {85}},\
  \bibinfo {pages} {2392} (\bibinfo {year} {2000})}\BibitemShut {NoStop}%
\bibitem [{\citenamefont {Blais}\ \emph {et~al.}(2004)\citenamefont {Blais},
  \citenamefont {Huang}, \citenamefont {Wallraff}, \citenamefont {Girvin},\
  and\ \citenamefont {Schoelkopf}}]{PhysRevA.69.062320}%
  \BibitemOpen
  \bibfield  {author} {\bibinfo {author} {\bibfnamefont {A.}~\bibnamefont
  {Blais}}, \bibinfo {author} {\bibfnamefont {R.-S.}\ \bibnamefont {Huang}},
  \bibinfo {author} {\bibfnamefont {A.}~\bibnamefont {Wallraff}}, \bibinfo
  {author} {\bibfnamefont {S.~M.}\ \bibnamefont {Girvin}},\ and\ \bibinfo
  {author} {\bibfnamefont {R.~J.}\ \bibnamefont {Schoelkopf}},\ }\bibfield
  {title} {\bibinfo {title} {Cavity quantum electrodynamics for superconducting
  electrical circuits: An architecture for quantum computation},\ }\href
  {https://doi.org/10.1103/PhysRevA.69.062320} {\bibfield  {journal} {\bibinfo
  {journal} {Phys. Rev. A}\ }\textbf {\bibinfo {volume} {69}},\ \bibinfo
  {pages} {062320} (\bibinfo {year} {2004})}\BibitemShut {NoStop}%
\bibitem [{\citenamefont {Blais}\ \emph {et~al.}(2021)\citenamefont {Blais},
  \citenamefont {Grimsmo}, \citenamefont {Girvin},\ and\ \citenamefont
  {Wallraff}}]{RevModPhys.93.025005}%
  \BibitemOpen
  \bibfield  {author} {\bibinfo {author} {\bibfnamefont {A.}~\bibnamefont
  {Blais}}, \bibinfo {author} {\bibfnamefont {A.~L.}\ \bibnamefont {Grimsmo}},
  \bibinfo {author} {\bibfnamefont {S.~M.}\ \bibnamefont {Girvin}},\ and\
  \bibinfo {author} {\bibfnamefont {A.}~\bibnamefont {Wallraff}},\ }\bibfield
  {title} {\bibinfo {title} {Circuit quantum electrodynamics},\ }\href
  {https://doi.org/10.1103/RevModPhys.93.025005} {\bibfield  {journal}
  {\bibinfo  {journal} {Rev. Mod. Phys.}\ }\textbf {\bibinfo {volume} {93}},\
  \bibinfo {pages} {025005} (\bibinfo {year} {2021})}\BibitemShut {NoStop}%
\bibitem [{\citenamefont {Petersson}\ \emph {et~al.}(2012)\citenamefont
  {Petersson}, \citenamefont {McFaul}, \citenamefont {Schroer}, \citenamefont
  {Jung}, \citenamefont {Taylor}, \citenamefont {Houck},\ and\ \citenamefont
  {Petta}}]{Petersson:2012aa}%
  \BibitemOpen
  \bibfield  {author} {\bibinfo {author} {\bibfnamefont {K.~D.}\ \bibnamefont
  {Petersson}}, \bibinfo {author} {\bibfnamefont {L.~W.}\ \bibnamefont
  {McFaul}}, \bibinfo {author} {\bibfnamefont {M.~D.}\ \bibnamefont {Schroer}},
  \bibinfo {author} {\bibfnamefont {M.}~\bibnamefont {Jung}}, \bibinfo {author}
  {\bibfnamefont {J.~M.}\ \bibnamefont {Taylor}}, \bibinfo {author}
  {\bibfnamefont {A.~A.}\ \bibnamefont {Houck}},\ and\ \bibinfo {author}
  {\bibfnamefont {J.~R.}\ \bibnamefont {Petta}},\ }\bibfield  {title} {\bibinfo
  {title} {Circuit quantum electrodynamics with a spin qubit},\ }\href
  {https://doi.org/10.1038/nature11559} {\bibfield  {journal} {\bibinfo
  {journal} {Nature}\ }\textbf {\bibinfo {volume} {490}},\ \bibinfo {pages}
  {380} (\bibinfo {year} {2012})}\BibitemShut {NoStop}%
\bibitem [{\citenamefont {Samkharadze}\ \emph {et~al.}(2018)\citenamefont
  {Samkharadze}, \citenamefont {Zheng}, \citenamefont {Kalhor}, \citenamefont
  {Brousse}, \citenamefont {Sammak}, \citenamefont {Mendes}, \citenamefont
  {Blais}, \citenamefont {Scappucci},\ and\ \citenamefont
  {Vandersypen}}]{doi:10.1126/science.aar4054}%
  \BibitemOpen
  \bibfield  {author} {\bibinfo {author} {\bibfnamefont {N.}~\bibnamefont
  {Samkharadze}}, \bibinfo {author} {\bibfnamefont {G.}~\bibnamefont {Zheng}},
  \bibinfo {author} {\bibfnamefont {N.}~\bibnamefont {Kalhor}}, \bibinfo
  {author} {\bibfnamefont {D.}~\bibnamefont {Brousse}}, \bibinfo {author}
  {\bibfnamefont {A.}~\bibnamefont {Sammak}}, \bibinfo {author} {\bibfnamefont
  {U.~C.}\ \bibnamefont {Mendes}}, \bibinfo {author} {\bibfnamefont
  {A.}~\bibnamefont {Blais}}, \bibinfo {author} {\bibfnamefont
  {G.}~\bibnamefont {Scappucci}},\ and\ \bibinfo {author} {\bibfnamefont
  {L.~M.~K.}\ \bibnamefont {Vandersypen}},\ }\bibfield  {title} {\bibinfo
  {title} {Strong spin-photon coupling in silicon},\ }\href
  {https://doi.org/10.1126/science.aar4054} {\bibfield  {journal} {\bibinfo
  {journal} {Science}\ }\textbf {\bibinfo {volume} {359}},\ \bibinfo {pages}
  {1123} (\bibinfo {year} {2018})}\BibitemShut {NoStop}%
\bibitem [{\citenamefont {Yu}\ \emph {et~al.}(2023)\citenamefont {Yu},
  \citenamefont {Zihlmann}, \citenamefont {Abadillo-Uriel}, \citenamefont
  {Michal}, \citenamefont {Rambal}, \citenamefont {Niebojewski}, \citenamefont
  {Bedecarrats}, \citenamefont {Vinet}, \citenamefont {Dumur}, \citenamefont
  {Filippone}, \citenamefont {Bertrand}, \citenamefont {De~Franceschi},
  \citenamefont {Niquet},\ and\ \citenamefont {Maurand}}]{Yu:2023aa}%
  \BibitemOpen
  \bibfield  {author} {\bibinfo {author} {\bibfnamefont {C.~X.}\ \bibnamefont
  {Yu}}, \bibinfo {author} {\bibfnamefont {S.}~\bibnamefont {Zihlmann}},
  \bibinfo {author} {\bibfnamefont {J.}~\bibnamefont {Abadillo-Uriel}},
  \bibinfo {author} {\bibfnamefont {V.~P.}\ \bibnamefont {Michal}}, \bibinfo
  {author} {\bibfnamefont {N.}~\bibnamefont {Rambal}}, \bibinfo {author}
  {\bibfnamefont {H.}~\bibnamefont {Niebojewski}}, \bibinfo {author}
  {\bibfnamefont {T.}~\bibnamefont {Bedecarrats}}, \bibinfo {author}
  {\bibfnamefont {M.}~\bibnamefont {Vinet}}, \bibinfo {author} {\bibfnamefont
  {{\'E}.}~\bibnamefont {Dumur}}, \bibinfo {author} {\bibfnamefont
  {M.}~\bibnamefont {Filippone}}, \bibinfo {author} {\bibfnamefont
  {B.}~\bibnamefont {Bertrand}}, \bibinfo {author} {\bibfnamefont
  {S.}~\bibnamefont {De~Franceschi}}, \bibinfo {author} {\bibfnamefont {Y.-M.}\
  \bibnamefont {Niquet}},\ and\ \bibinfo {author} {\bibfnamefont
  {R.}~\bibnamefont {Maurand}},\ }\bibfield  {title} {\bibinfo {title} {Strong
  coupling between a photon and a hole spin in silicon},\ }\href
  {https://doi.org/10.1038/s41565-023-01332-3} {\bibfield  {journal} {\bibinfo
  {journal} {Nature Nanotechnology}\ }\textbf {\bibinfo {volume} {18}},\
  \bibinfo {pages} {741} (\bibinfo {year} {2023})}\BibitemShut {NoStop}%
\bibitem [{\citenamefont {B{\o}ttcher}\ \emph {et~al.}(2022)\citenamefont
  {B{\o}ttcher}, \citenamefont {Harvey}, \citenamefont {Fallahi}, \citenamefont
  {Gardner}, \citenamefont {Manfra}, \citenamefont {Vool}, \citenamefont
  {Bartlett},\ and\ \citenamefont {Yacoby}}]{Bottcher:2022aa}%
  \BibitemOpen
  \bibfield  {author} {\bibinfo {author} {\bibfnamefont {C.~G.~L.}\
  \bibnamefont {B{\o}ttcher}}, \bibinfo {author} {\bibfnamefont {S.~P.}\
  \bibnamefont {Harvey}}, \bibinfo {author} {\bibfnamefont {S.}~\bibnamefont
  {Fallahi}}, \bibinfo {author} {\bibfnamefont {G.~C.}\ \bibnamefont
  {Gardner}}, \bibinfo {author} {\bibfnamefont {M.~J.}\ \bibnamefont {Manfra}},
  \bibinfo {author} {\bibfnamefont {U.}~\bibnamefont {Vool}}, \bibinfo {author}
  {\bibfnamefont {S.~D.}\ \bibnamefont {Bartlett}},\ and\ \bibinfo {author}
  {\bibfnamefont {A.}~\bibnamefont {Yacoby}},\ }\bibfield  {title} {\bibinfo
  {title} {Parametric longitudinal coupling between a high-impedance
  superconducting resonator and a semiconductor quantum dot singlet-triplet
  spin qubit},\ }\href {https://doi.org/10.1038/s41467-022-32236-w} {\bibfield
  {journal} {\bibinfo  {journal} {Nature Communications}\ }\textbf {\bibinfo
  {volume} {13}},\ \bibinfo {pages} {4773} (\bibinfo {year}
  {2022})}\BibitemShut {NoStop}%
\bibitem [{\citenamefont {Corrigan}\ \emph {et~al.}(2023)\citenamefont
  {Corrigan}, \citenamefont {Harpt}, \citenamefont {Holman}, \citenamefont
  {Ruskov}, \citenamefont {Marciniec}, \citenamefont {Rosenberg}, \citenamefont
  {Yost}, \citenamefont {Das}, \citenamefont {Oliver}, \citenamefont
  {McDermott}, \citenamefont {Tahan}, \citenamefont {Friesen},\ and\
  \citenamefont {Eriksson}}]{PhysRevApplied.20.064005}%
  \BibitemOpen
  \bibfield  {author} {\bibinfo {author} {\bibfnamefont {J.}~\bibnamefont
  {Corrigan}}, \bibinfo {author} {\bibfnamefont {B.}~\bibnamefont {Harpt}},
  \bibinfo {author} {\bibfnamefont {N.}~\bibnamefont {Holman}}, \bibinfo
  {author} {\bibfnamefont {R.}~\bibnamefont {Ruskov}}, \bibinfo {author}
  {\bibfnamefont {P.}~\bibnamefont {Marciniec}}, \bibinfo {author}
  {\bibfnamefont {D.}~\bibnamefont {Rosenberg}}, \bibinfo {author}
  {\bibfnamefont {D.}~\bibnamefont {Yost}}, \bibinfo {author} {\bibfnamefont
  {R.}~\bibnamefont {Das}}, \bibinfo {author} {\bibfnamefont {W.~D.}\
  \bibnamefont {Oliver}}, \bibinfo {author} {\bibfnamefont {R.}~\bibnamefont
  {McDermott}}, \bibinfo {author} {\bibfnamefont {C.}~\bibnamefont {Tahan}},
  \bibinfo {author} {\bibfnamefont {M.}~\bibnamefont {Friesen}},\ and\ \bibinfo
  {author} {\bibfnamefont {M.}~\bibnamefont {Eriksson}},\ }\bibfield  {title}
  {\bibinfo {title} {Longitudinal coupling between a
  ${\mathrm{si}/\mathrm{si}}_{1\ensuremath{-}x}{\mathrm{ge}}_{x}$ double
  quantum dot and an off-chip $\mathrm{Ti}\mathrm{N}$ resonator},\ }\href
  {https://doi.org/10.1103/PhysRevApplied.20.064005} {\bibfield  {journal}
  {\bibinfo  {journal} {Phys. Rev. Appl.}\ }\textbf {\bibinfo {volume} {20}},\
  \bibinfo {pages} {064005} (\bibinfo {year} {2023})}\BibitemShut {NoStop}%
\bibitem [{\citenamefont {Li}\ \emph {et~al.}(2018{\natexlab{b}})\citenamefont
  {Li}, \citenamefont {Liu}, \citenamefont {Wu},\ and\ \citenamefont
  {Liu}}]{RL2018b}%
  \BibitemOpen
  \bibfield  {author} {\bibinfo {author} {\bibfnamefont {R.}~\bibnamefont
  {Li}}, \bibinfo {author} {\bibfnamefont {Z.~H.}\ \bibnamefont {Liu}},
  \bibinfo {author} {\bibfnamefont {Y.}~\bibnamefont {Wu}},\ and\ \bibinfo
  {author} {\bibfnamefont {C.~S.}\ \bibnamefont {Liu}},\ }\bibfield  {title}
  {\bibinfo {title} {The impacts of the quantum-dot confining potential on the
  spin-orbit effect},\ }\href {https://doi.org/10.1038/s41598-018-25692-2}
  {\bibfield  {journal} {\bibinfo  {journal} {Sci Rep}\ }\textbf {\bibinfo
  {volume} {8}},\ \bibinfo {pages} {7400} (\bibinfo {year}
  {2018}{\natexlab{b}})}\BibitemShut {NoStop}%
\bibitem [{\citenamefont {Ban}\ \emph {et~al.}(2015)\citenamefont {Ban},
  \citenamefont {Chen}, \citenamefont {Muga},\ and\ \citenamefont
  {Sherman}}]{PhysRevA.91.023604}%
  \BibitemOpen
  \bibfield  {author} {\bibinfo {author} {\bibfnamefont {Y.}~\bibnamefont
  {Ban}}, \bibinfo {author} {\bibfnamefont {X.}~\bibnamefont {Chen}}, \bibinfo
  {author} {\bibfnamefont {J.~G.}\ \bibnamefont {Muga}},\ and\ \bibinfo
  {author} {\bibfnamefont {E.~Y.}\ \bibnamefont {Sherman}},\ }\bibfield
  {title} {\bibinfo {title} {Quantum state engineering of spin-orbit-coupled
  ultracold atoms in a morse potential},\ }\href
  {https://doi.org/10.1103/PhysRevA.91.023604} {\bibfield  {journal} {\bibinfo
  {journal} {Phys. Rev. A}\ }\textbf {\bibinfo {volume} {91}},\ \bibinfo
  {pages} {023604} (\bibinfo {year} {2015})}\BibitemShut {NoStop}%
\bibitem [{\citenamefont {Nadj-Perge}\ \emph {et~al.}(2012)\citenamefont
  {Nadj-Perge}, \citenamefont {Pribiag}, \citenamefont {van~den Berg},
  \citenamefont {Zuo}, \citenamefont {Plissard}, \citenamefont {Bakkers},
  \citenamefont {Frolov},\ and\ \citenamefont
  {Kouwenhoven}}]{nadj2012spectroscopy}%
  \BibitemOpen
  \bibfield  {author} {\bibinfo {author} {\bibfnamefont {S.}~\bibnamefont
  {Nadj-Perge}}, \bibinfo {author} {\bibfnamefont {V.~S.}\ \bibnamefont
  {Pribiag}}, \bibinfo {author} {\bibfnamefont {J.~W.~G.}\ \bibnamefont
  {van~den Berg}}, \bibinfo {author} {\bibfnamefont {K.}~\bibnamefont {Zuo}},
  \bibinfo {author} {\bibfnamefont {S.~R.}\ \bibnamefont {Plissard}}, \bibinfo
  {author} {\bibfnamefont {E.~P. A.~M.}\ \bibnamefont {Bakkers}}, \bibinfo
  {author} {\bibfnamefont {S.~M.}\ \bibnamefont {Frolov}},\ and\ \bibinfo
  {author} {\bibfnamefont {L.~P.}\ \bibnamefont {Kouwenhoven}},\ }\bibfield
  {title} {\bibinfo {title} {Spectroscopy of spin-orbit quantum bits in indium
  antimonide nanowires},\ }\href
  {https://doi.org/10.1103/PhysRevLett.108.166801} {\bibfield  {journal}
  {\bibinfo  {journal} {Phys. Rev. Lett.}\ }\textbf {\bibinfo {volume} {108}},\
  \bibinfo {pages} {166801} (\bibinfo {year} {2012})}\BibitemShut {NoStop}%
\bibitem [{\citenamefont {van~den Berg}\ \emph {et~al.}(2013)\citenamefont
  {van~den Berg}, \citenamefont {Nadj-Perge}, \citenamefont {Pribiag},
  \citenamefont {Plissard}, \citenamefont {Bakkers}, \citenamefont {Frolov},\
  and\ \citenamefont {Kouwenhoven}}]{PhysRevLett.110.066806}%
  \BibitemOpen
  \bibfield  {author} {\bibinfo {author} {\bibfnamefont {J.~W.~G.}\
  \bibnamefont {van~den Berg}}, \bibinfo {author} {\bibfnamefont
  {S.}~\bibnamefont {Nadj-Perge}}, \bibinfo {author} {\bibfnamefont {V.~S.}\
  \bibnamefont {Pribiag}}, \bibinfo {author} {\bibfnamefont {S.~R.}\
  \bibnamefont {Plissard}}, \bibinfo {author} {\bibfnamefont {E.~P. A.~M.}\
  \bibnamefont {Bakkers}}, \bibinfo {author} {\bibfnamefont {S.~M.}\
  \bibnamefont {Frolov}},\ and\ \bibinfo {author} {\bibfnamefont {L.~P.}\
  \bibnamefont {Kouwenhoven}},\ }\bibfield  {title} {\bibinfo {title} {Fast
  spin-orbit qubit in an indium antimonide nanowire},\ }\href
  {https://doi.org/10.1103/PhysRevLett.110.066806} {\bibfield  {journal}
  {\bibinfo  {journal} {Phys. Rev. Lett.}\ }\textbf {\bibinfo {volume} {110}},\
  \bibinfo {pages} {066806} (\bibinfo {year} {2013})}\BibitemShut {NoStop}%
\bibitem [{\citenamefont {Froning}\ \emph {et~al.}(2021)\citenamefont
  {Froning}, \citenamefont {Camenzind}, \citenamefont {van~der Molen},
  \citenamefont {Li}, \citenamefont {Bakkers}, \citenamefont {Zumb{\"u}hl},\
  and\ \citenamefont {Braakman}}]{Froning:2021aa}%
  \BibitemOpen
  \bibfield  {author} {\bibinfo {author} {\bibfnamefont {F.~N.~M.}\
  \bibnamefont {Froning}}, \bibinfo {author} {\bibfnamefont {L.~C.}\
  \bibnamefont {Camenzind}}, \bibinfo {author} {\bibfnamefont {O.~A.~H.}\
  \bibnamefont {van~der Molen}}, \bibinfo {author} {\bibfnamefont
  {A.}~\bibnamefont {Li}}, \bibinfo {author} {\bibfnamefont {E.~P. A.~M.}\
  \bibnamefont {Bakkers}}, \bibinfo {author} {\bibfnamefont {D.~M.}\
  \bibnamefont {Zumb{\"u}hl}},\ and\ \bibinfo {author} {\bibfnamefont {F.~R.}\
  \bibnamefont {Braakman}},\ }\bibfield  {title} {\bibinfo {title} {Ultrafast
  hole spin qubit with gate-tunable spin--orbit switch functionality},\ }\href
  {https://doi.org/10.1038/s41565-020-00828-6} {\bibfield  {journal} {\bibinfo
  {journal} {Nature Nanotechnology}\ }\textbf {\bibinfo {volume} {16}},\
  \bibinfo {pages} {308} (\bibinfo {year} {2021})}\BibitemShut {NoStop}%
\bibitem [{\citenamefont {Wang}\ \emph {et~al.}(2022)\citenamefont {Wang},
  \citenamefont {Xu}, \citenamefont {Gao}, \citenamefont {Liu}, \citenamefont
  {Ma}, \citenamefont {Zhang}, \citenamefont {Wang}, \citenamefont {Cao},
  \citenamefont {Wang}, \citenamefont {Zhang}, \citenamefont {Culcer},
  \citenamefont {Hu}, \citenamefont {Jiang}, \citenamefont {Li}, \citenamefont
  {Guo},\ and\ \citenamefont {Guo}}]{Wang:2022tm}%
  \BibitemOpen
  \bibfield  {author} {\bibinfo {author} {\bibfnamefont {K.}~\bibnamefont
  {Wang}}, \bibinfo {author} {\bibfnamefont {G.}~\bibnamefont {Xu}}, \bibinfo
  {author} {\bibfnamefont {F.}~\bibnamefont {Gao}}, \bibinfo {author}
  {\bibfnamefont {H.}~\bibnamefont {Liu}}, \bibinfo {author} {\bibfnamefont
  {R.-L.}\ \bibnamefont {Ma}}, \bibinfo {author} {\bibfnamefont
  {X.}~\bibnamefont {Zhang}}, \bibinfo {author} {\bibfnamefont
  {Z.}~\bibnamefont {Wang}}, \bibinfo {author} {\bibfnamefont {G.}~\bibnamefont
  {Cao}}, \bibinfo {author} {\bibfnamefont {T.}~\bibnamefont {Wang}}, \bibinfo
  {author} {\bibfnamefont {J.-J.}\ \bibnamefont {Zhang}}, \bibinfo {author}
  {\bibfnamefont {D.}~\bibnamefont {Culcer}}, \bibinfo {author} {\bibfnamefont
  {X.}~\bibnamefont {Hu}}, \bibinfo {author} {\bibfnamefont {H.-W.}\
  \bibnamefont {Jiang}}, \bibinfo {author} {\bibfnamefont {H.-O.}\ \bibnamefont
  {Li}}, \bibinfo {author} {\bibfnamefont {G.-C.}\ \bibnamefont {Guo}},\ and\
  \bibinfo {author} {\bibfnamefont {G.-P.}\ \bibnamefont {Guo}},\ }\bibfield
  {title} {\bibinfo {title} {Ultrafast coherent control of a hole spin qubit in
  a germanium quantum dot},\ }\href
  {https://doi.org/10.1038/s41467-021-27880-7} {\bibfield  {journal} {\bibinfo
  {journal} {Nature Communications}\ }\textbf {\bibinfo {volume} {13}},\
  \bibinfo {pages} {206} (\bibinfo {year} {2022})}\BibitemShut {NoStop}%
\bibitem [{\citenamefont {Gambetta}\ \emph {et~al.}(2015)\citenamefont
  {Gambetta}, \citenamefont {Ziani}, \citenamefont {Barbarino}, \citenamefont
  {Cavaliere},\ and\ \citenamefont {Sassetti}}]{PhysRevB.91.235421}%
  \BibitemOpen
  \bibfield  {author} {\bibinfo {author} {\bibfnamefont {F.~M.}\ \bibnamefont
  {Gambetta}}, \bibinfo {author} {\bibfnamefont {N.~T.}\ \bibnamefont {Ziani}},
  \bibinfo {author} {\bibfnamefont {S.}~\bibnamefont {Barbarino}}, \bibinfo
  {author} {\bibfnamefont {F.}~\bibnamefont {Cavaliere}},\ and\ \bibinfo
  {author} {\bibfnamefont {M.}~\bibnamefont {Sassetti}},\ }\bibfield  {title}
  {\bibinfo {title} {Anomalous friedel oscillations in a quasihelical quantum
  dot},\ }\href {https://doi.org/10.1103/PhysRevB.91.235421} {\bibfield
  {journal} {\bibinfo  {journal} {Phys. Rev. B}\ }\textbf {\bibinfo {volume}
  {91}},\ \bibinfo {pages} {235421} (\bibinfo {year} {2015})}\BibitemShut
  {NoStop}%
\bibitem [{\citenamefont {Madelung}(2004)}]{madelung2004semiconductors}%
  \BibitemOpen
  \bibfield  {author} {\bibinfo {author} {\bibfnamefont {O.}~\bibnamefont
  {Madelung}},\ }\href@noop {} {\emph {\bibinfo {title} {Semiconductors: data
  handbook}}}\ (\bibinfo  {publisher} {Springer Science \& Business Media},\
  \bibinfo {year} {2004})\BibitemShut {NoStop}%
\bibitem [{\citenamefont {Bychkov}\ and\ \citenamefont
  {Rashba}(1984)}]{bychkov1984oscillatory}%
  \BibitemOpen
  \bibfield  {author} {\bibinfo {author} {\bibfnamefont {Y.~A.}\ \bibnamefont
  {Bychkov}}\ and\ \bibinfo {author} {\bibfnamefont {E.~I.}\ \bibnamefont
  {Rashba}},\ }\bibfield  {title} {\bibinfo {title} {Oscillatory effects and
  the magnetic susceptibility of carriers in inversion layers},\ }\href
  {https://doi.org/10.1088/0022-3719/17/33/015} {\bibfield  {journal} {\bibinfo
   {journal} {Journal of Physics C: Solid State Physics}\ }\textbf {\bibinfo
  {volume} {17}},\ \bibinfo {pages} {6039} (\bibinfo {year}
  {1984})}\BibitemShut {NoStop}%
\bibitem [{\citenamefont {Tsitsishvili}\ \emph {et~al.}(2004)\citenamefont
  {Tsitsishvili}, \citenamefont {Lozano},\ and\ \citenamefont
  {Gogolin}}]{PhysRevB.70.115316}%
  \BibitemOpen
  \bibfield  {author} {\bibinfo {author} {\bibfnamefont {E.}~\bibnamefont
  {Tsitsishvili}}, \bibinfo {author} {\bibfnamefont {G.~S.}\ \bibnamefont
  {Lozano}},\ and\ \bibinfo {author} {\bibfnamefont {A.~O.}\ \bibnamefont
  {Gogolin}},\ }\bibfield  {title} {\bibinfo {title} {Rashba coupling in
  quantum dots: An exact solution},\ }\href
  {https://doi.org/10.1103/PhysRevB.70.115316} {\bibfield  {journal} {\bibinfo
  {journal} {Phys. Rev. B}\ }\textbf {\bibinfo {volume} {70}},\ \bibinfo
  {pages} {115316} (\bibinfo {year} {2004})}\BibitemShut {NoStop}%
\bibitem [{\citenamefont {Dresselhaus}(1955)}]{PhysRev.100.580}%
  \BibitemOpen
  \bibfield  {author} {\bibinfo {author} {\bibfnamefont {G.}~\bibnamefont
  {Dresselhaus}},\ }\bibfield  {title} {\bibinfo {title} {Spin-orbit coupling
  effects in zinc blende structures},\ }\href
  {https://doi.org/10.1103/PhysRev.100.580} {\bibfield  {journal} {\bibinfo
  {journal} {Phys. Rev.}\ }\textbf {\bibinfo {volume} {100}},\ \bibinfo {pages}
  {580} (\bibinfo {year} {1955})}\BibitemShut {NoStop}%
\bibitem [{\citenamefont {Nowak}\ and\ \citenamefont
  {Szafran}(2013)}]{PhysRevB.87.205436}%
  \BibitemOpen
  \bibfield  {author} {\bibinfo {author} {\bibfnamefont {M.~P.}\ \bibnamefont
  {Nowak}}\ and\ \bibinfo {author} {\bibfnamefont {B.}~\bibnamefont
  {Szafran}},\ }\bibfield  {title} {\bibinfo {title} {Spin-polarization
  anisotropy in a narrow spin-orbit-coupled nanowire quantum dot},\ }\href
  {https://doi.org/10.1103/PhysRevB.87.205436} {\bibfield  {journal} {\bibinfo
  {journal} {Phys. Rev. B}\ }\textbf {\bibinfo {volume} {87}},\ \bibinfo
  {pages} {205436} (\bibinfo {year} {2013})}\BibitemShut {NoStop}%
\bibitem [{\citenamefont {Kloeffel}\ \emph {et~al.}(2011)\citenamefont
  {Kloeffel}, \citenamefont {Trif},\ and\ \citenamefont
  {Loss}}]{PhysRevB.84.195314}%
  \BibitemOpen
  \bibfield  {author} {\bibinfo {author} {\bibfnamefont {C.}~\bibnamefont
  {Kloeffel}}, \bibinfo {author} {\bibfnamefont {M.}~\bibnamefont {Trif}},\
  and\ \bibinfo {author} {\bibfnamefont {D.}~\bibnamefont {Loss}},\ }\bibfield
  {title} {\bibinfo {title} {Strong spin-orbit interaction and helical hole
  states in {Ge/Si} nanowires},\ }\href
  {https://doi.org/10.1103/PhysRevB.84.195314} {\bibfield  {journal} {\bibinfo
  {journal} {Phys. Rev. B}\ }\textbf {\bibinfo {volume} {84}},\ \bibinfo
  {pages} {195314} (\bibinfo {year} {2011})}\BibitemShut {NoStop}%
\bibitem [{\citenamefont {Li}(2022)}]{RL2022a}%
  \BibitemOpen
  \bibfield  {author} {\bibinfo {author} {\bibfnamefont {R.}~\bibnamefont
  {Li}},\ }\bibfield  {title} {\bibinfo {title} {Searching strong `spin'-orbit
  coupled one-dimensional hole gas in strong magnetic fields},\ }\href
  {https://doi.org/10.1088/1361-648x/ac37da} {\bibfield  {journal} {\bibinfo
  {journal} {Journal of Physics: Condensed Matter}\ }\textbf {\bibinfo {volume}
  {34}},\ \bibinfo {pages} {075301} (\bibinfo {year} {2022})}\BibitemShut
  {NoStop}%
\bibitem [{\citenamefont {Li}\ and\ \citenamefont {Qi}(2023)}]{RL2023b}%
  \BibitemOpen
  \bibfield  {author} {\bibinfo {author} {\bibfnamefont {R.}~\bibnamefont
  {Li}}\ and\ \bibinfo {author} {\bibfnamefont {X.-Y.}\ \bibnamefont {Qi}},\
  }\bibfield  {title} {\bibinfo {title} {Two-band description of the strong
  `spin'-orbit coupled one-dimensional hole gas in a cylindrical {Ge}
  nanowire},\ }\bibfield  {booktitle} {\emph {\bibinfo {booktitle} {Journal of
  Physics: Condensed Matter}},\ }\href
  {https://doi.org/10.1088/1361-648X/acb8f5} {\ \textbf {\bibinfo {volume}
  {35}},\ \bibinfo {pages} {135302} (\bibinfo {year} {2023})}\BibitemShut
  {NoStop}%
\bibitem [{\citenamefont {Li}\ and\ \citenamefont {Li}(2023)}]{RL2023c}%
  \BibitemOpen
  \bibfield  {author} {\bibinfo {author} {\bibfnamefont {R.}~\bibnamefont
  {Li}}\ and\ \bibinfo {author} {\bibfnamefont {Z.-Q.}\ \bibnamefont {Li}},\
  }\bibfield  {title} {\bibinfo {title} {Low-energy hole subband dispersions in
  a cylindrical {Ge} nanowire: the effects of the nanowire growth direction},\
  }\bibfield  {booktitle} {\emph {\bibinfo {booktitle} {Journal of Physics:
  Condensed Matter}},\ }\href {https://doi.org/10.1088/1361-648X/acd674} {\
  \textbf {\bibinfo {volume} {35}},\ \bibinfo {pages} {345301} (\bibinfo {year}
  {2023})}\BibitemShut {NoStop}%
\bibitem [{\citenamefont {Kloeffel}\ \emph {et~al.}(2018)\citenamefont
  {Kloeffel}, \citenamefont {Ran\ifmmode \check{c}\else
  \v{c}\fi{}i\ifmmode~\acute{c}\else \'{c}\fi{}},\ and\ \citenamefont
  {Loss}}]{PhysRevB.97.235422}%
  \BibitemOpen
  \bibfield  {author} {\bibinfo {author} {\bibfnamefont {C.}~\bibnamefont
  {Kloeffel}}, \bibinfo {author} {\bibfnamefont {M.~J.}\ \bibnamefont
  {Ran\ifmmode \check{c}\else \v{c}\fi{}i\ifmmode~\acute{c}\else \'{c}\fi{}}},\
  and\ \bibinfo {author} {\bibfnamefont {D.}~\bibnamefont {Loss}},\ }\bibfield
  {title} {\bibinfo {title} {Direct rashba spin-orbit interaction in {Si} and
  {Ge} nanowires with different growth directions},\ }\href
  {https://doi.org/10.1103/PhysRevB.97.235422} {\bibfield  {journal} {\bibinfo
  {journal} {Phys. Rev. B}\ }\textbf {\bibinfo {volume} {97}},\ \bibinfo
  {pages} {235422} (\bibinfo {year} {2018})}\BibitemShut {NoStop}%
\bibitem [{\citenamefont {Luo}\ \emph {et~al.}(2011)\citenamefont {Luo},
  \citenamefont {Zhang},\ and\ \citenamefont {Zunger}}]{PhysRevB.84.121303}%
  \BibitemOpen
  \bibfield  {author} {\bibinfo {author} {\bibfnamefont {J.-W.}\ \bibnamefont
  {Luo}}, \bibinfo {author} {\bibfnamefont {L.}~\bibnamefont {Zhang}},\ and\
  \bibinfo {author} {\bibfnamefont {A.}~\bibnamefont {Zunger}},\ }\bibfield
  {title} {\bibinfo {title} {Absence of intrinsic spin splitting in
  one-dimensional quantum wires of tetrahedral semiconductors},\ }\href
  {https://doi.org/10.1103/PhysRevB.84.121303} {\bibfield  {journal} {\bibinfo
  {journal} {Phys. Rev. B}\ }\textbf {\bibinfo {volume} {84}},\ \bibinfo
  {pages} {121303} (\bibinfo {year} {2011})}\BibitemShut {NoStop}%
\bibitem [{\citenamefont {Luo}\ \emph {et~al.}(2017)\citenamefont {Luo},
  \citenamefont {Li},\ and\ \citenamefont {Zunger}}]{PhysRevLett.119.126401}%
  \BibitemOpen
  \bibfield  {author} {\bibinfo {author} {\bibfnamefont {J.-W.}\ \bibnamefont
  {Luo}}, \bibinfo {author} {\bibfnamefont {S.-S.}\ \bibnamefont {Li}},\ and\
  \bibinfo {author} {\bibfnamefont {A.}~\bibnamefont {Zunger}},\ }\bibfield
  {title} {\bibinfo {title} {Rapid transition of the hole rashba effect from
  strong field dependence to saturation in semiconductor nanowires},\ }\href
  {https://doi.org/10.1103/PhysRevLett.119.126401} {\bibfield  {journal}
  {\bibinfo  {journal} {Phys. Rev. Lett.}\ }\textbf {\bibinfo {volume} {119}},\
  \bibinfo {pages} {126401} (\bibinfo {year} {2017})}\BibitemShut {NoStop}%
\bibitem [{\citenamefont {Maier}\ \emph {et~al.}(2014)\citenamefont {Maier},
  \citenamefont {Klinovaja},\ and\ \citenamefont {Loss}}]{PhysRevB.90.195421}%
  \BibitemOpen
  \bibfield  {author} {\bibinfo {author} {\bibfnamefont {F.}~\bibnamefont
  {Maier}}, \bibinfo {author} {\bibfnamefont {J.}~\bibnamefont {Klinovaja}},\
  and\ \bibinfo {author} {\bibfnamefont {D.}~\bibnamefont {Loss}},\ }\bibfield
  {title} {\bibinfo {title} {Majorana fermions in {Ge/Si} hole nanowires},\
  }\href {https://doi.org/10.1103/PhysRevB.90.195421} {\bibfield  {journal}
  {\bibinfo  {journal} {Phys. Rev. B}\ }\textbf {\bibinfo {volume} {90}},\
  \bibinfo {pages} {195421} (\bibinfo {year} {2014})}\BibitemShut {NoStop}%
\bibitem [{\citenamefont {Milivojevi\ifmmode~\acute{c}\else
  \'{c}\fi{}}(2021)}]{PhysRevB.104.235304}%
  \BibitemOpen
  \bibfield  {author} {\bibinfo {author} {\bibfnamefont {M.}~\bibnamefont
  {Milivojevi\ifmmode~\acute{c}\else \'{c}\fi{}}},\ }\bibfield  {title}
  {\bibinfo {title} {Electrical control of the hole spin qubit in {S}i and {G}e
  nanowire quantum dots},\ }\href {https://doi.org/10.1103/PhysRevB.104.235304}
  {\bibfield  {journal} {\bibinfo  {journal} {Phys. Rev. B}\ }\textbf {\bibinfo
  {volume} {104}},\ \bibinfo {pages} {235304} (\bibinfo {year}
  {2021})}\BibitemShut {NoStop}%
\bibitem [{\citenamefont {Adelsberger}\ \emph
  {et~al.}(2022{\natexlab{a}})\citenamefont {Adelsberger}, \citenamefont
  {Benito}, \citenamefont {Bosco}, \citenamefont {Klinovaja},\ and\
  \citenamefont {Loss}}]{PhysRevB.105.075308}%
  \BibitemOpen
  \bibfield  {author} {\bibinfo {author} {\bibfnamefont {C.}~\bibnamefont
  {Adelsberger}}, \bibinfo {author} {\bibfnamefont {M.}~\bibnamefont {Benito}},
  \bibinfo {author} {\bibfnamefont {S.}~\bibnamefont {Bosco}}, \bibinfo
  {author} {\bibfnamefont {J.}~\bibnamefont {Klinovaja}},\ and\ \bibinfo
  {author} {\bibfnamefont {D.}~\bibnamefont {Loss}},\ }\bibfield  {title}
  {\bibinfo {title} {Hole-spin qubits in {Ge} nanowire quantum dots: Interplay
  of orbital magnetic field, strain, and growth direction},\ }\href
  {https://doi.org/10.1103/PhysRevB.105.075308} {\bibfield  {journal} {\bibinfo
   {journal} {Phys. Rev. B}\ }\textbf {\bibinfo {volume} {105}},\ \bibinfo
  {pages} {075308} (\bibinfo {year} {2022}{\natexlab{a}})}\BibitemShut
  {NoStop}%
\bibitem [{\citenamefont {Adelsberger}\ \emph
  {et~al.}(2022{\natexlab{b}})\citenamefont {Adelsberger}, \citenamefont
  {Bosco}, \citenamefont {Klinovaja},\ and\ \citenamefont
  {Loss}}]{PhysRevB.106.235408}%
  \BibitemOpen
  \bibfield  {author} {\bibinfo {author} {\bibfnamefont {C.}~\bibnamefont
  {Adelsberger}}, \bibinfo {author} {\bibfnamefont {S.}~\bibnamefont {Bosco}},
  \bibinfo {author} {\bibfnamefont {J.}~\bibnamefont {Klinovaja}},\ and\
  \bibinfo {author} {\bibfnamefont {D.}~\bibnamefont {Loss}},\ }\bibfield
  {title} {\bibinfo {title} {Enhanced orbital magnetic field effects in {Ge}
  hole nanowires},\ }\href {https://doi.org/10.1103/PhysRevB.106.235408}
  {\bibfield  {journal} {\bibinfo  {journal} {Phys. Rev. B}\ }\textbf {\bibinfo
  {volume} {106}},\ \bibinfo {pages} {235408} (\bibinfo {year}
  {2022}{\natexlab{b}})}\BibitemShut {NoStop}%
\bibitem [{\citenamefont {Buluta}\ \emph {et~al.}(2011)\citenamefont {Buluta},
  \citenamefont {Ashhab},\ and\ \citenamefont {Nori}}]{Buluta_2011}%
  \BibitemOpen
  \bibfield  {author} {\bibinfo {author} {\bibfnamefont {I.}~\bibnamefont
  {Buluta}}, \bibinfo {author} {\bibfnamefont {S.}~\bibnamefont {Ashhab}},\
  and\ \bibinfo {author} {\bibfnamefont {F.}~\bibnamefont {Nori}},\ }\bibfield
  {title} {\bibinfo {title} {Natural and artificial atoms for quantum
  computation},\ }\href {https://doi.org/10.1088/0034-4885/74/10/104401}
  {\bibfield  {journal} {\bibinfo  {journal} {Reports on Progress in Physics}\
  }\textbf {\bibinfo {volume} {74}},\ \bibinfo {pages} {104401} (\bibinfo
  {year} {2011})}\BibitemShut {NoStop}%
\bibitem [{\citenamefont {Paladino}\ \emph {et~al.}(2014)\citenamefont
  {Paladino}, \citenamefont {Galperin}, \citenamefont {Falci},\ and\
  \citenamefont {Altshuler}}]{RevModPhys.86.361}%
  \BibitemOpen
  \bibfield  {author} {\bibinfo {author} {\bibfnamefont {E.}~\bibnamefont
  {Paladino}}, \bibinfo {author} {\bibfnamefont {Y.~M.}\ \bibnamefont
  {Galperin}}, \bibinfo {author} {\bibfnamefont {G.}~\bibnamefont {Falci}},\
  and\ \bibinfo {author} {\bibfnamefont {B.~L.}\ \bibnamefont {Altshuler}},\
  }\bibfield  {title} {\bibinfo {title} {$1/f$ noise: Implications for
  solid-state quantum information},\ }\href
  {https://doi.org/10.1103/RevModPhys.86.361} {\bibfield  {journal} {\bibinfo
  {journal} {Rev. Mod. Phys.}\ }\textbf {\bibinfo {volume} {86}},\ \bibinfo
  {pages} {361} (\bibinfo {year} {2014})}\BibitemShut {NoStop}%
\bibitem [{\citenamefont {Hayashi}\ \emph {et~al.}(2003)\citenamefont
  {Hayashi}, \citenamefont {Fujisawa}, \citenamefont {Cheong}, \citenamefont
  {Jeong},\ and\ \citenamefont {Hirayama}}]{PhysRevLett.91.226804}%
  \BibitemOpen
  \bibfield  {author} {\bibinfo {author} {\bibfnamefont {T.}~\bibnamefont
  {Hayashi}}, \bibinfo {author} {\bibfnamefont {T.}~\bibnamefont {Fujisawa}},
  \bibinfo {author} {\bibfnamefont {H.~D.}\ \bibnamefont {Cheong}}, \bibinfo
  {author} {\bibfnamefont {Y.~H.}\ \bibnamefont {Jeong}},\ and\ \bibinfo
  {author} {\bibfnamefont {Y.}~\bibnamefont {Hirayama}},\ }\bibfield  {title}
  {\bibinfo {title} {Coherent manipulation of electronic states in a double
  quantum dot},\ }\href {https://doi.org/10.1103/PhysRevLett.91.226804}
  {\bibfield  {journal} {\bibinfo  {journal} {Phys. Rev. Lett.}\ }\textbf
  {\bibinfo {volume} {91}},\ \bibinfo {pages} {226804} (\bibinfo {year}
  {2003})}\BibitemShut {NoStop}%
\bibitem [{\citenamefont {Gorman}\ \emph {et~al.}(2005)\citenamefont {Gorman},
  \citenamefont {Hasko},\ and\ \citenamefont
  {Williams}}]{PhysRevLett.95.090502}%
  \BibitemOpen
  \bibfield  {author} {\bibinfo {author} {\bibfnamefont {J.}~\bibnamefont
  {Gorman}}, \bibinfo {author} {\bibfnamefont {D.~G.}\ \bibnamefont {Hasko}},\
  and\ \bibinfo {author} {\bibfnamefont {D.~A.}\ \bibnamefont {Williams}},\
  }\bibfield  {title} {\bibinfo {title} {Charge-qubit operation of an isolated
  double quantum dot},\ }\href {https://doi.org/10.1103/PhysRevLett.95.090502}
  {\bibfield  {journal} {\bibinfo  {journal} {Phys. Rev. Lett.}\ }\textbf
  {\bibinfo {volume} {95}},\ \bibinfo {pages} {090502} (\bibinfo {year}
  {2005})}\BibitemShut {NoStop}%
\bibitem [{\citenamefont {Li}(2019)}]{RL2019}%
  \BibitemOpen
  \bibfield  {author} {\bibinfo {author} {\bibfnamefont {R.}~\bibnamefont
  {Li}},\ }\bibfield  {title} {\bibinfo {title} {Spin manipulation and spin
  dephasing in quantum dot integrated with a slanting magnetic field},\ }\href
  {https://doi.org/10.1088/1402-4896/ab1358} {\bibfield  {journal} {\bibinfo
  {journal} {Physica Scripta}\ }\textbf {\bibinfo {volume} {94}},\ \bibinfo
  {pages} {085808} (\bibinfo {year} {2019})}\BibitemShut {NoStop}%
\end{thebibliography}%
\end{document}